\documentclass[a4paper,12pt]{article}
\usepackage{test} 
\usepackage[authoryear,round,longnamesfirst]{natbib}

\usepackage[inline,shortlabels]{enumitem}
\setlist[enumerate,1]{label=(\roman*)}

\usepackage{color}
\usepackage[colorlinks,citecolor=blue,urlcolor=magenta]{hyperref}
\usepackage{doi}
\usepackage{xurl} 
\usepackage{booktabs} 
\usepackage{caption,subcaption}
\usepackage{graphicx}

\usepackage[margin = 1.25in]{geometry}
\usepackage{setspace}
\allowdisplaybreaks

\usepackage{bbm}
\renewcommand{\hat}{\widehat} 
\renewcommand{\Pr}{\operatorname{P}}

\title{Capital and Labor Income Pareto Exponents in the United States, 1916--2019}

\author{\normalsize
Ji Hyung Lee\thanks{\footnotesize\setlength{\baselineskip}{4.4mm} 
Ji Hyung Lee: \href{mailto:jihyung@illinois.edu}{jihyung@illinois.edu}. Department of Economics, University of Illinois, 214 David Kinley Hall, 1407 West Gregory Drive, Urbana, IL 61801, USA}
\and\normalsize
Yuya Sasaki\thanks{\footnotesize\setlength{\baselineskip}{4.4mm} Yuya Sasaki: \href{mailto:yuya.sasaki@vanderbilt.edu}{yuya.sasaki@vanderbilt.edu}. Department of Economics, Vanderbilt University, VU Station B \#351819, 2301 Vanderbilt Place, Nashville, TN 37235-1819, USA\smallskip}
\and\normalsize
Alexis Akira Toda\thanks{\footnotesize\setlength{\baselineskip}{4.4mm} Alexis Akira Toda: \href{mailto:atoda@ucsd.edu}{atoda@ucsd.edu}. Department of Economics, University of California San Diego, 9500 Gilman Dr, \#0508, La Jolla, CA 92093-0508, USA\smallskip}
\and\normalsize
Yulong Wang\thanks{\footnotesize\setlength{\baselineskip}{4.4mm} Yulong Wang: \href{mailto:ywang402@syr.edu}{ywang402@syr.edu}. Department of Economics, Syracuse University, 110 Eggers Hall, Syracuse, NY 13244-1020, USA}
}

\date{\today}

\numberwithin{equation}{section}
\numberwithin{thm}{section}

\begin{document}
\maketitle

\begin{abstract}

Accurately estimating income Pareto exponents is challenging due to limitations in data availability and the applicability of statistical methods. Using tabulated summaries of incomes from tax authorities and a recent estimation method, we estimate income Pareto exponents in U.S. for 1916-2019. We find that during the past three decades, the capital and labor income Pareto exponents have been stable at around 1.2 and 2. Our findings suggest that the top tail income and wealth inequality is higher and wealthy agents have twice as large an impact on the aggregate economy than previously thought but there is no clear trend post-1985.

\medskip


\medskip

\end{abstract}

\section{Introduction}

The Pareto exponent characterizes the heaviness of the upper tail of size distributions.\footnote{We say that a positive random variable $X$ obeys a power law with Pareto exponent $\alpha>0$ if the tail probability decays like a power function: $\Pr(X>x)\sim x^{-\alpha}$ for large $x$.} The income Pareto exponent is not only a convenient descriptive statistic for top tail inequality but also plays an important role in economic theory. One example is the theory of optimal taxation. When the government's objective is to maximize the social welfare subject to raising a target tax revenue, \cite{Saez2001} shows that the optimal top marginal income tax rate is $\tau=\frac{1-g}{1-g+\alpha e}$, where $g\ge 0$ is the utility weight on top income earners, $\alpha$ is the income Pareto exponent, and $e$ is the elasticity of top incomes with respect to the tax rate. This tax rate is clearly highly dependent on the Pareto exponent $\alpha$, and any policy recommendation based on economic theory requires an accurate estimate of $\alpha$. Another example is the calibration of quantitative macroeconomic models. When the income distribution has a Pareto upper tail with exponent $\alpha>1$, the average income of individuals above a certain income threshold is $\frac{\alpha}{\alpha-1}$ times the income threshold. This multiple is highly dependent on $\alpha$ and large when $\alpha$ is close to 1 (Zipf's law). Getting the magnitude of $\alpha$ right is important for the calibration of quantitative macroeconomic models because the rich have a substantial impact on aggregate quantities.\footnote{See \cite{BeareToda2022ECMA} for how to determine Pareto exponents in dynamic economic models and \cite{Gouin-BonenfantToda2022QE} for how to numerically solve such models when the wealth distribution is Pareto-tailed. On the empirical side, \cite{TodaWalsh2020RFS} document that the top income share negatively predicts stock market returns.}

Despite its importance, accurately estimating income Pareto exponents is challenging due to limitations in data availability and the applicability of statistical methods. Survey data on income such as the \emph{Current Population Survey}, \emph{Panel Study of Income Dynamics}, and \emph{Survey of Consumer Finances} are not necessarily reliable due to low response rates or inaccurate responses. Furthermore, the sample sizes are typically no more than a few thousands, which makes it difficult to grasp the behavior of rich households. Although administrative data such as tax returns have much larger sample sizes (well over a million) and responses are likely more accurate (because income earners above some thresholds are legally required to file for taxes and are subject to the possibility of audits), individual-level data are either difficult to access\footnote{As of the time of writing, Internal Revenue Service (IRS) releases Public Use Tax File that samples 0.07\% of the tax returns only for years 2009--2014 and charges the user about \$10,000 for accessing the data. Furthermore, to protect the identity of individuals, the income amounts are statistically altered, for example by adding a random noise.} or nonexistent for the historically old periods and only tabulated summaries are publicly available. Turning to estimation, commonly applied statistical methods such as maximum likelihood \citep{hill1975}, log rank regression \citep{GabaixIbragimov2011}, and other methods\footnote{See \cite{GomesGuillou2015} and \cite{Fedotenkov2020} for reviews of estimation methods for Pareto exponents.} often require individual-level data. 
Although the maximum likelihood method is applicable with grouped data if income thresholds that define the income groups and the number of individuals in each group are observable, in practice the income thresholds are observable only for total income and not for capital or labor income, making it impossible to estimate capital and labor income Pareto exponents separately. This is undesirable because capital and labor income are taxed differently and an understanding of their separate tail behavior is crucial for policy recommendations.

This paper overcomes all the challenges just described and estimates the capital and labor income Pareto exponents in the United States for the period 1916--2019. Regarding data availability and quality, we use only tabulated summaries of tax returns, which are publicly available for over a century and the sample sizes range from a few to over a hundred million, which are more than enough to accurately estimate the Pareto exponents. For estimation, we apply the recently developed method of \cite{TodaWang2021JAE}, which is based on extreme value theory and maximum likelihood. The main advantages of this method are that
\begin{enumerate*}
\item it takes into account the dependence between order statistics,
\item it does not require the observation of thresholds that define income groups, which are available only for adjusted gross income, and
\item it only requires the observation of group averages (or top income shares), which can be easily recovered from tabulated summary data.
\end{enumerate*}

We find that the sample period (early twentieth century to present) can be divided into three sub periods based on Pareto exponents, pre-1940, 1940--1985, and post-1985. The capital income Pareto exponent is roughly constant at 1.2 pre-1940 and post-1985, and shows an inverse U-shaped pattern during 1940--1985, plateauing at 2.0 in 1960--1970. The total and labor income Pareto exponents show a qualitatively similar pattern but the magnitude is larger (top tail inequality is lower), with pre-1940 and post-1985 point estimates of around 1.5 and 2, respectively.

Relative to the existing literature, our paper contributes to a deeper understanding of top tail income inequality in several dimensions. First, to our knowledge, our paper is the first to systematically report long time series of income Pareto exponents using the best publicly available data and the best statistical method. As we review shortly, existing papers either consider only a long time series of total income (without distinguishing capital and labor income due to the limitations in estimation methods) or provide analyses of separate income categories but only for particular years (due to data availability). Second, our point estimates of Pareto exponents tend to be smaller than those reported in the existing literature, which may force researchers to update their prior beliefs about the magnitude of the Pareto exponents. For instance, in an influential review article, \citet[p.~275]{gabaix2009} mentions that ``the tail exponent of [total] income seems to vary between 1.5 and 3'' (citing \citealp{AtkinsonPiketty2007}) and ``it seems that the tail exponent of wealth is rather stable, perhaps around 1.5'' (citing \citealp{KlassBihamLevyMalcaiSolomon2006}). The conjectured number 1.5 for wealth is supported by the more recent study of \citet[Table 8]{Vermeulen2018}, who combines data from \emph{Survey of Consumer Finances} and \emph{Forbes 400} to estimate the wealth Pareto exponent. Although we do not observe wealth from income tax returns data, if we are willing to assume that capital income is proportional to wealth, then our results suggest that the wealth exponent assumes a lower value of 1.2, closer to Zipf's law. The fact that our estimate is smaller than the ones in the existing literature suggests that survey or Forbes data suffer from either underreporting or mismeasurement of intangible wealth. As we mentioned at the beginning of the introduction, since the impact of wealthy agents on aggregate quantities is proportional to $\frac{\alpha}{\alpha-1}$, the impact doubles from $1.5/0.5=3$ with the previous estimate to $1.2/0.2=6$ with the new one. Finally, our Pareto exponent estimates are remarkably stable since 1985 and do not show particular trends. This may be surprising because it is widely documented that income inequality measured by top income shares has risen exactly in this period \citep{PikettySaez2003}. Our results suggest that the rise in income inequality during the past three or four decades concerns the inequality between the rich and the rest, and not the top tail inequality.


\subsection*{Related literature}

\cite{Pareto1895LaLegge,Pareto1896LaCourbe,Pareto1897Cours} used tabulated summary data of tax returns in many European states to document that the tail probability of the income distribution exhibits a straight-line pattern in a log-log plot (or the tail probability decays like a power function), which is now called Pareto's law or the power law. In particular, \cite{Pareto1895LaLegge} assumes the probability density function $f(y)=Hy^{-h}$ (so the Pareto exponent is $\alpha=h-1$) and estimates $h$ by simple curve fitting. He finds that the Pareto exponent $\alpha$ is in a narrow range of 1.3--1.7 (see the table on p.~61). More recent research that employs micro data include \cite{reed2001}, \cite{Toda2012JEBO}, and \cite{SaezStancheva2018} for U.S., \cite{reed2003} for U.S., Canada, Sri Lanka, and Bohemia, \cite{nirei-souma2007} for Japan, \cite{Jenkins2017} for U.K., and \cite{IbragimovIbragimov2018} for Russia. These papers all concern specific countries and years and thus are not systematic.

\citet[Table A-1]{FeenbergPoterba1993} estimate the Pareto exponents of adjusted gross income (AGI) in U.S. for the period 1951--1990. Using the cumulative distribution function (CDF) $F(y)=1-ky^{-\alpha}$ for the Pareto distribution, they find two income thresholds $y_1<y_2$ that bracket the top 0.5\% of taxpayers and estimate $\alpha$ as
\begin{equation}
\hat{\alpha}=\log[(1-F(y_1))/(1-F(y_2))]/\log(y_2/y_1), \label{eq:alpha_FP}
\end{equation}
where $F(y_1)$ and $F(y_2)$ are computed from the tabulated summary. The estimated exponent is roughly single-peaked, increasing from 1.83 in 1951 to 2.46 in 1970 and then decreasing to 1.59 in 1990.

When income has a Pareto upper tail with exponent $\alpha>1$, it is straightforward to show that the top $p$ fractile income share $S(p)$ is proportional to $p^{1-1/\alpha}$ (see \eqref{eq:Sp} below). Using this property, \citet[Appendix B, Section 1.1, pp. 592–599]{Piketty2001book} infers the local Pareto exponent between income thresholds in the tabulated summary and interpolates top income shares for arbitrary top fractile $p$. Applying this method, \citet[Table 13A.23]{AtkinsonPiketty2010} report the Pareto exponent estimated as
\begin{equation}
\hat{\alpha}=\left(1-\frac{\log[S(q)/S(p)]}{\log(q/p)}\right)^{-1} \label{eq:alpha_AP}
\end{equation}
for $(p,q)=(0.001,0.01)$ for many countries and years. However, the methods described above are all heuristic and lack rigorous statistical foundations.

\cite{deVriesTodaLIS} estimate the capital and labor income Pareto exponents across 475 country-year observations from the \emph{Luxembourg Income Study} (LIS) that span 52 countries over half a century (1967–2018) and document that capital income inequality is higher than labor income inequality (median Pareto exponents 1.46 and 3.35 respectively). As LIS is a database of individual survey data, it has limitations in sample size and accuracy. In fact, as we document in Figure \ref{fig:alpha_LIS}, their estimates tend to be biased upwards (underestimate inequality) possibly due to underreporting by rich households.

\section{Data}\label{sec:data}

\subsection{General framework}\label{subsec:data_gen}

Consider the latent sample $\set{Y_i}_{i=1}^n$ of the values $Y_i$ of income, which is not directly observed by the researcher. Here $i$ indexes tax units and $n$ is the sample size. We denote the order statistics in descending order by
\begin{equation*}
Y_{(1)}\ge Y_{(2)}\ge \dots\ge Y_{(n)}.
\end{equation*}
For each $m\le n$, define the partial sum of order statistics
\begin{equation}
S_m\coloneqq \sum_{i=1}^m Y_{(i)}. \label{eq:partial_sum}
\end{equation}
Publicly available data on the income distribution released from tax authorities often take the form of tabulations of partial sums $\set{S_{n_k}}_{k=1}^K$, where $n_1<\dots<n_K\le n$ is an increasing sequence of natural numbers (the number of tax filers included in the partial sum \eqref{eq:partial_sum}) and $K$ denotes the number of income groups. Our problem is to estimate the Pareto exponent of the income distribution based only on the cumulative number of top income earners $\set{n_k}_{k=1}^K$ and the partial sums of top incomes $\set{S_{n_k}}_{k=1}^K$.

\subsection{U.S. income data}\label{subsec:data_us}

Our primary data source is the \emph{Statistics of Income (SOI) Individual Income Tax Returns Publication 1304} from the United States Internal Revenue Service. For each year we record the thresholds for adjusted gross income (AGI) as well as the number of returns and the total income accruing to each group, both for ``Adjusted gross income less deficit'' and ``Salaries and wages''. See Appendix \ref{sec:detail} for details on the data collecting procedure. Table \ref{t:IRS2019} presents an example data set from the 2019 U.S.\ tax returns.

\begin{table}[!htb]
\centering
\caption{Income distribution in the United States, 2019.}\label{t:IRS2019}
\begin{tabular}{rrrrrr}
\toprule
\multicolumn2{c}{Income group} & \multicolumn2{c}{Adjusted gross income (AGI)} & \multicolumn2{c}{Salaries and wages} \\
& \multicolumn1{c}{(1)} & \multicolumn1{c}{(2)} & \multicolumn1{c}{(3)} & \multicolumn1{c}{(4)} & \multicolumn1{c}{(5)} \\
$k$ & AGI threshold & \# returns & Total income & \# returns & Total income \\
\cmidrule(lr){1-2}
\cmidrule(lr){3-4}
\cmidrule(lr){5-6}
& $-\infty$ & 2,127,500 & -237,064,231 & 569,047 & 23,421,857 \\
18 & \$1 &	9,866,880	&	24,439,988	&	6,672,531	&	23,927,191	\\
17 & \$5,000 &	9,925,940	&	74,584,857	&	7,622,306	&	58,927,624	\\
16 & \$10,000 &	11,087,737	&	138,230,399	&	8,277,447	&	100,631,554	\\
15 & \$15,000 &	10,039,446	&	175,255,963	&	7,931,946	&	134,897,400	\\
14 & \$20,000 &	9,493,968	&	213,660,160	&	7,855,283	&	173,142,941	\\
13 & \$25,000 &	9,289,939	&	254,877,708	&	7,943,835	&	212,428,275	\\
12 & \$30,000 &	16,090,602	&	560,073,192	&	14,045,867	&	471,544,226	\\
11 & \$40,000 &	12,503,041	&	560,258,808	&	10,931,707	&	465,547,848	\\
10 & \$50,000 &	22,238,948	&	1,366,892,948	&	18,976,338	&	1,071,062,478	\\
9 & \$75,000 &	14,118,568	&	1,222,947,425	&	12,033,727	&	921,390,540	\\
8 & \$100,000 &	21,997,582	&	3,004,363,636	&	19,028,674	&	2,209,484,837	\\
7 & \$200,000 &	7,297,883	&	2,090,808,696	&	6,414,121	&	1,429,162,189	\\
6 & \$500,000 &	1,162,371	&	781,920,814	&	1,010,488	&	449,489,139	\\
5 & \$1,000,000 &	254,197	&	305,561,848	&	214,955	&	141,101,999	\\
4 & \$1,500,000 &	103,075	&	176,961,208	&	85,285	&	72,754,006	\\
3 & \$2,000,000 &	143,514	&	425,088,995	&	117,168	&	145,270,762	\\
2 & \$5,000,000 &	34,738	&	237,781,553	&	28,162	&	66,367,353	\\
1 & \$10,000,000 &	20,876	&	590,230,011	&	16,866	&	102,518,828	\\
\cmidrule(lr){1-2}
\cmidrule(lr){3-4}
\cmidrule(lr){5-6}
\multicolumn2{c}{All returns, total} & 157,796,807 & 11,966,873,976 & 129,775,754 & 8,273,071,046 \\
\bottomrule
\end{tabular}
\caption*{\footnotesize Note: ``AGI threshold'' is the lower threshold of adjusted gross income (AGI) that defines the income groups. ``\# returns'' is the number of tax returns with income weakly above the lower AGI threshold and strictly below the upper AGI threshold. ``Total income'' is the total income (in units of 1,000 U.S.\ dollars) accruing to tax filers in each income group. The row ``All returns, total'' shows the total for each column.}
\end{table}

In Table \ref{t:IRS2019}, the number of income groups is $K=18$. Column (1) shows the lower threshold of adjusted gross income (AGI) for each income group. Column (2) shows the number of tax filers within each income group, which corresponds to $n_k-n_{k-1}$ in our notation (where we set $n_0=0$ by convention). Column (3) shows the total income (AGI) accruing to tax filers in each income group in units of 1,000 U.S.\ dollars, which corresponds to $(S_{n_k}-S_{n_{k-1}})/1{,}000$ in our notation (where we set $S_0=0$ by convention). Thus for AGI, in addition to the cumulative number of top income earners $\set{n_k}_{k=1}^K$ and the partial sums of top incomes $\set{S_{n_k}}_{k=1}^K$, we also observe the income thresholds that define the income groups.

The U.S.\ tax returns data also contain information on other income categories such as ``salaries and wages'', ``taxable interest'', ``ordinary dividends'', etc. Columns (4) and (5) of Table \ref{t:IRS2019} show the number of tax returns and total income (salaries and wages) for each income group defined by AGI. Note that the numbers of returns in Column (4) are smaller than those in Column (2), because some tax filers reported no salaries or wages (possibly because they are business owners or retired). For example, among the 20{,}876 tax filers with adjusted gross income above \$10{,}000{,}000 (group 1), only 16{,}866 reported salaries and wages, implying that $20{,}876-16{,}866=4{,}010$ tax filers in this income group reported no salaries or wages. Furthermore, unlike the case with AGI, for salaries and wages, we do not observe the income thresholds for each group.

\subsection{Capital income}\label{subsec:defn_capital}

We are interested in estimating the Pareto exponents of the total, capital, and labor income. We define total and labor income as AGI and ``Salaries and wages'' in Table \ref{t:IRS2019}, respectively.

Unlike total and labor income, there is no clear-cut definition of capital income. One possibility is to interpret capital income broadly and simply define it as non-labor income (AGI minus salaries and wages). Another possibility is to add up income categories that can reasonably be classified as capital income, for instance interests, dividends, capital gains, etc. The advantage of the first approach is that it is simple and can be consistently applied throughout the sample period. The disadvantage is that it may be inaccurate because it contains incomes from clearly non-capital sources such as state income tax refunds, alimony received, and unemployment compensation. The second approach has the advantage that it is likely more accurate, but it could be ambiguous whether to include or exclude particular income categories as capital income because the tax legislation has significantly changed over the sample period. For instance, in 1918 there were only seven income categories: ``wages and salaries'', ``business'', ``partnerships, personal service corporations, estates, and trusts'', ``profits from sales of real estate, stocks, bonds, etc.'', ``rents and royalties'', ``dividends'', and ``interest and investment income''. The titles of these income categories suggest that the first (AGI minus salaries and wages) and second (adding up capital income components) approach would result in the same number (ignoring deductions). However, in 2019 there were 30 income categories and 16 statutory adjustments, which will likely make the capital income constructed using the two approaches different.

To address this issue, we compare the two definitions of capital income in 2019. The first approach simply defines capital income as non-labor income, which is AGI minus salaries and wages. For the second approach, we include the following categories as capital income: ``taxable interest'', ``tax-exempt interest'', ``ordinary dividends'', ``qualified dividends'', ``business or profession'', ``capital gain distributions reported on Form 1040'', ``sales of capital assets reported on Form 1040, Schedule D'', ``sales of property other than capital assets'', ``taxable Individual Retirement Arrangement (IRA) distributions'', ``pensions and annuities'', ``total rent and royalty'', ``partnership and S corporation'', and ``estate and trust''.\footnote{We categorize ``taxable Individual Retirement Arrangement (IRA) distributions'' and ``pensions and annuities'' as capital income because these incomes are likely derived from long-term investments. See Figure \ref{fig:average_income2} in Appendix \ref{subsec:robust_capital} for a robustness check that excludes these items.}

\begin{figure}[!htb]
\centering
\includegraphics[width=0.7\linewidth]{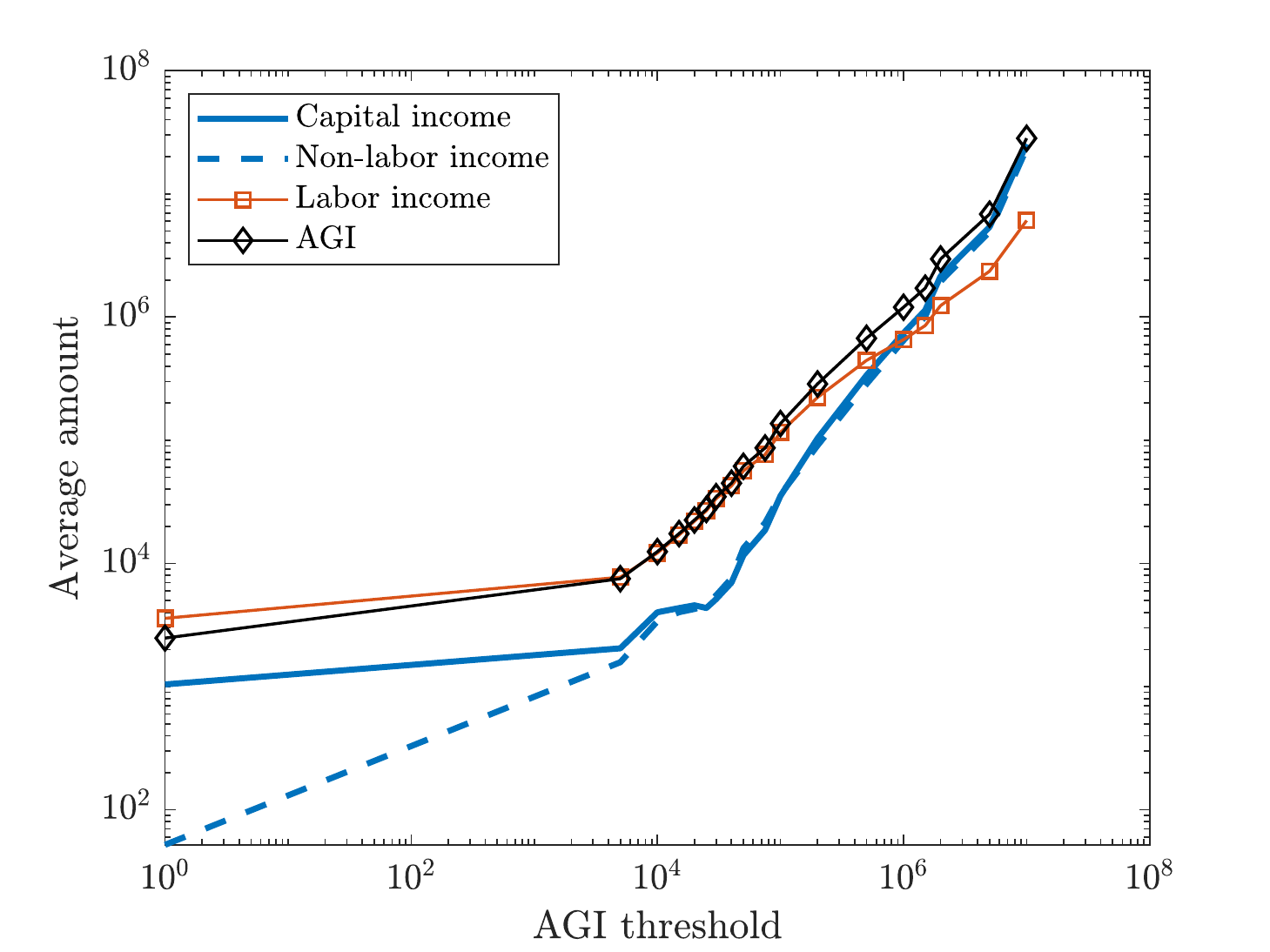}
\caption{Average income of each group, 2019.}\label{fig:average_income}
\caption*{\footnotesize Note: The figure shows the average income of each income group defined by the lower threshold of AGI. ``Capital income'' is defined by adding up capital income components. ``Non-labor income'' is AGI minus labor income (salaries and wages).}
\end{figure}

Figure \ref{fig:average_income} shows the average income of each group for capital income (second approach), non-labor income (first approach), labor income, and AGI. (The denominator of the average is Column (4) of Table \ref{t:IRS2019} for labor income and Column (2) for other incomes.) Figure \ref{fig:average_income} reveals several patterns. First, for income groups with AGI below \$100,000, the average labor income and AGI are nearly identical, which suggests that labor income is the primary income source for the middle class. Second, for income groups with AGI above \$1,000,000, the average capital income, non-labor income, and AGI are nearly identical, which suggests that capital income is the primary income source for the rich. Finally, and most importantly, for income groups with AGI above \$25,000, the capital incomes constructed from the two approaches (adding-up capital income components or subtracting labor income from AGI) are nearly identical. This finding suggests that it does not matter which approach we use to estimate the capital income Pareto exponent because we only use the top income groups.\footnote{As we discuss in Section \ref{sec:results}, we only use the top 1\% income groups for estimation, which correspond to tax filers with AGI above \$500,000 for 2019.} The intuition for this result is that for sufficiently rich tax filers, non-capital income other than salaries and wages (such as state income tax refunds, alimony received, unemployment compensation, and social security benefits) are relatively unimportant. In our subsequent empirical analysis, we use the first approach and define capital income as non-labor income (AGI minus salaries and wages) because it is simple and can be consistently applied throughout the sample period.

\begin{figure}[!htb]
\centering
\includegraphics[width=0.7\linewidth]{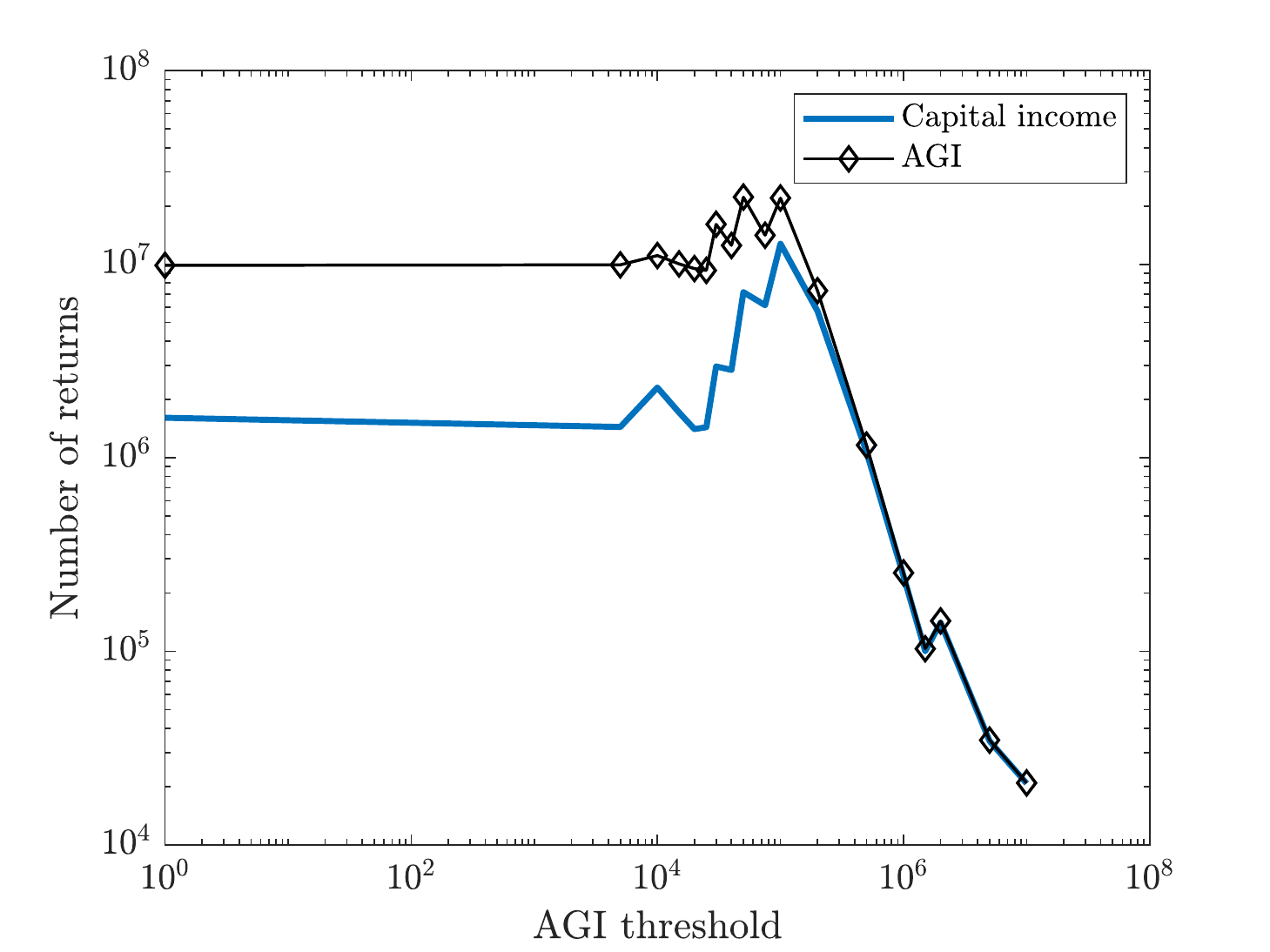}
\caption{Number of tax returns for AGI and capital income, 2019.}\label{fig:N_capital}
\caption*{\footnotesize Note: The figure shows the number of tax returns in each income group defined by the lower threshold of AGI. ``AGI'' and ``Capital income'' are the overall number of returns and those with positive capital income, respectively.}
\end{figure}

As we define capital income as AGI minus salaries and wages, we do not observe the cumulative number of top income earners $\set{n_k}_{k=1}^K$ for capital income. However, based on our argument that capital income is nearly identical to AGI for the rich (Figure \ref{fig:average_income}), we use the cumulative number of top income earners $\set{n_k}_{k=1}^K$ for AGI as those for capital income. To justify this choice, for 2019 we construct the number of tax filers with at least one capital income category included in the adding-up approach described above. Figure \ref{fig:N_capital} shows the number of tax returns and those with positive capital income for each income group. The two numbers are nearly identical above AGI \$200,000, which suggests that nearly all rich tax filers have positive capital income and hence the number of tax filers with positive capital income is nearly identical to the number of overall tax filers, at least for the rich.

\subsection{Sample size}\label{subsec:sample_size}

The unit of our analysis is the tax unit, which is either an individual or a married couple filing jointly (with dependents if any). However, we only observe the actual tax filers and the sample size $n$ (the universe of potential tax filers) is unobserved due to the presence of non-filers (\eg, those with income below the filing requirement or those who work in informal sectors using cash and evade taxes). Although the exact value of $n$ is unimportant because the Pareto distribution is scale invariant and the standard errors are minuscule due to the large sample size (at least ten million), the order of magnitude of $n$ matters because it affects the top fractiles used in the estimation: see \eqref{eq:pk}.

\begin{figure}[!htb]
\centering
\begin{subfigure}{0.48\linewidth}
\includegraphics[width=\linewidth]{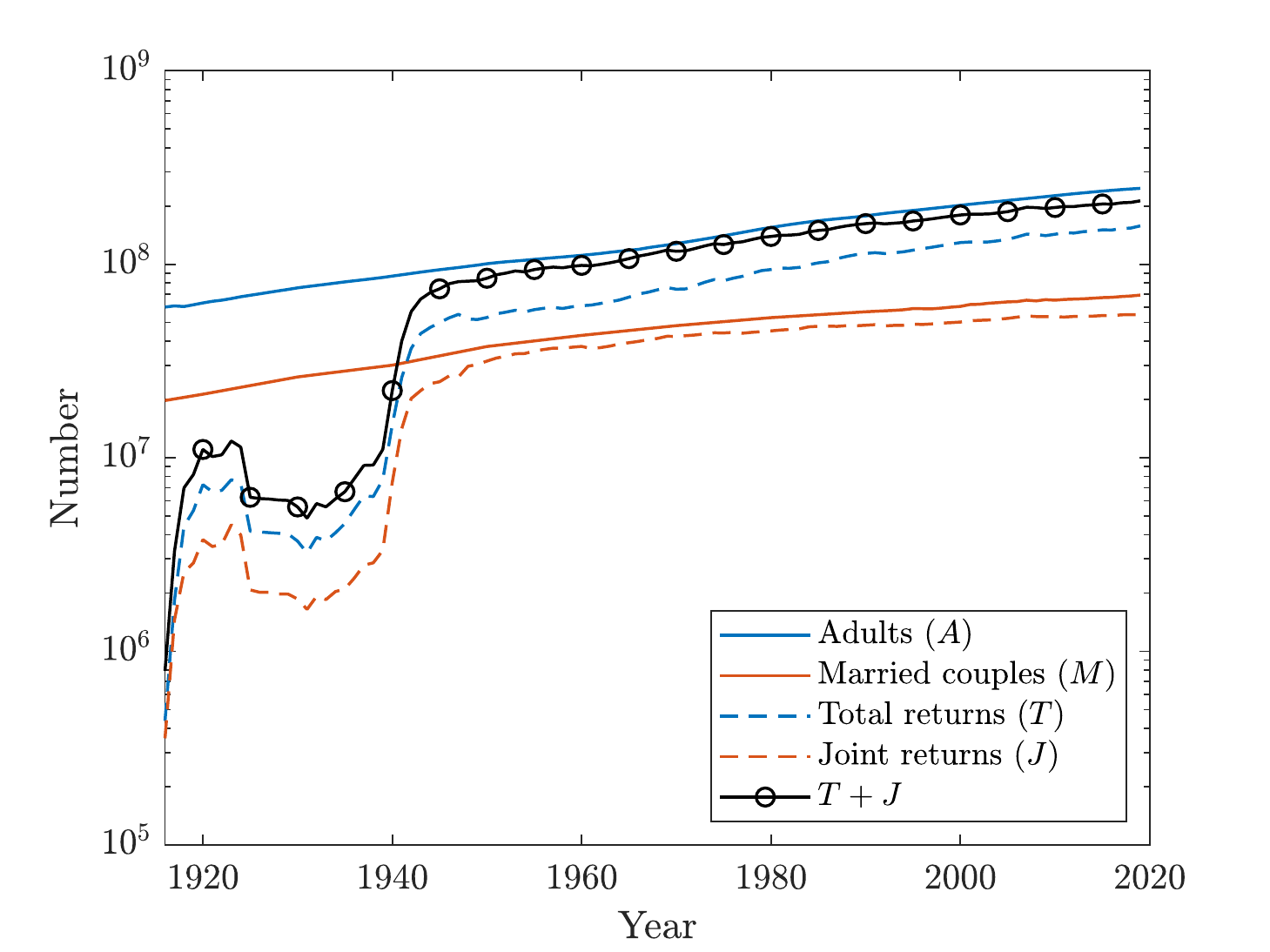}
\caption{Number of adults and tax returns.}\label{fig:adult}
\end{subfigure}
\begin{subfigure}{0.48\linewidth}
\includegraphics[width=\linewidth]{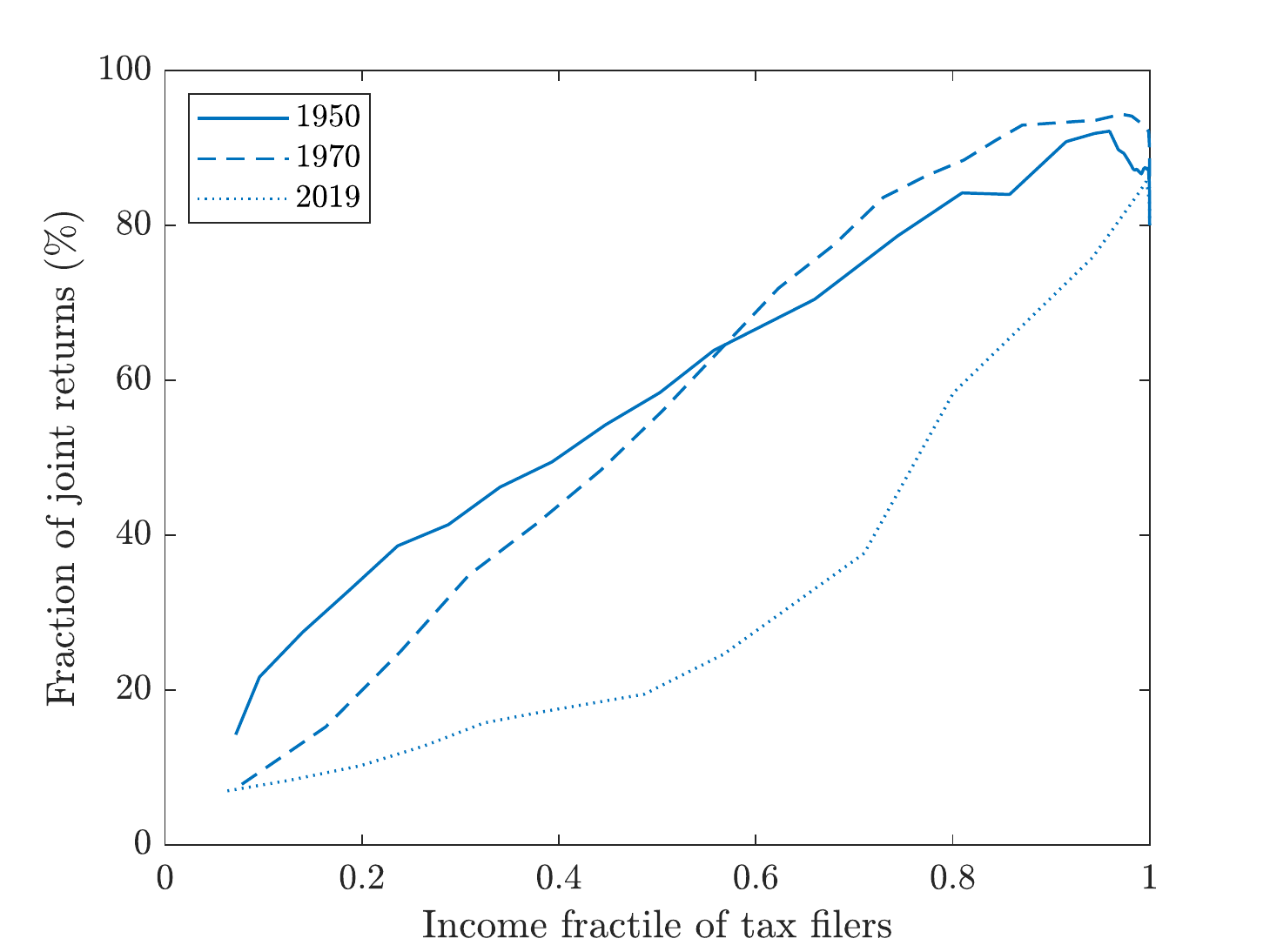}
\caption{Income fractile and joint returns.}\label{fig:frac_joint_fractile}
\end{subfigure}
\begin{subfigure}{0.48\linewidth}
\includegraphics[width=\linewidth]{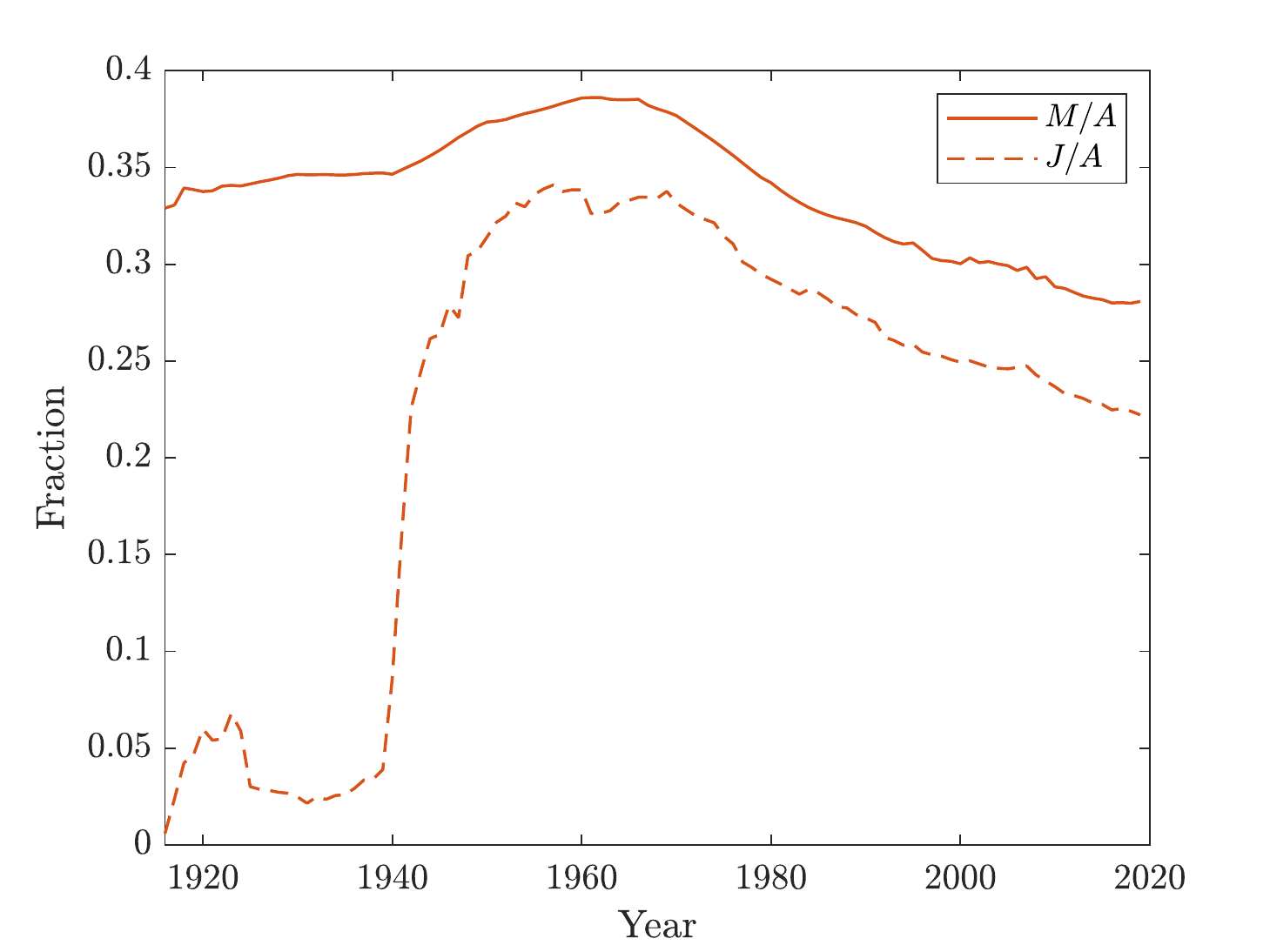}
\caption{Married couples and joint returns.}\label{fig:frac_MJ}
\end{subfigure}
\begin{subfigure}{0.48\linewidth}
\includegraphics[width=\linewidth]{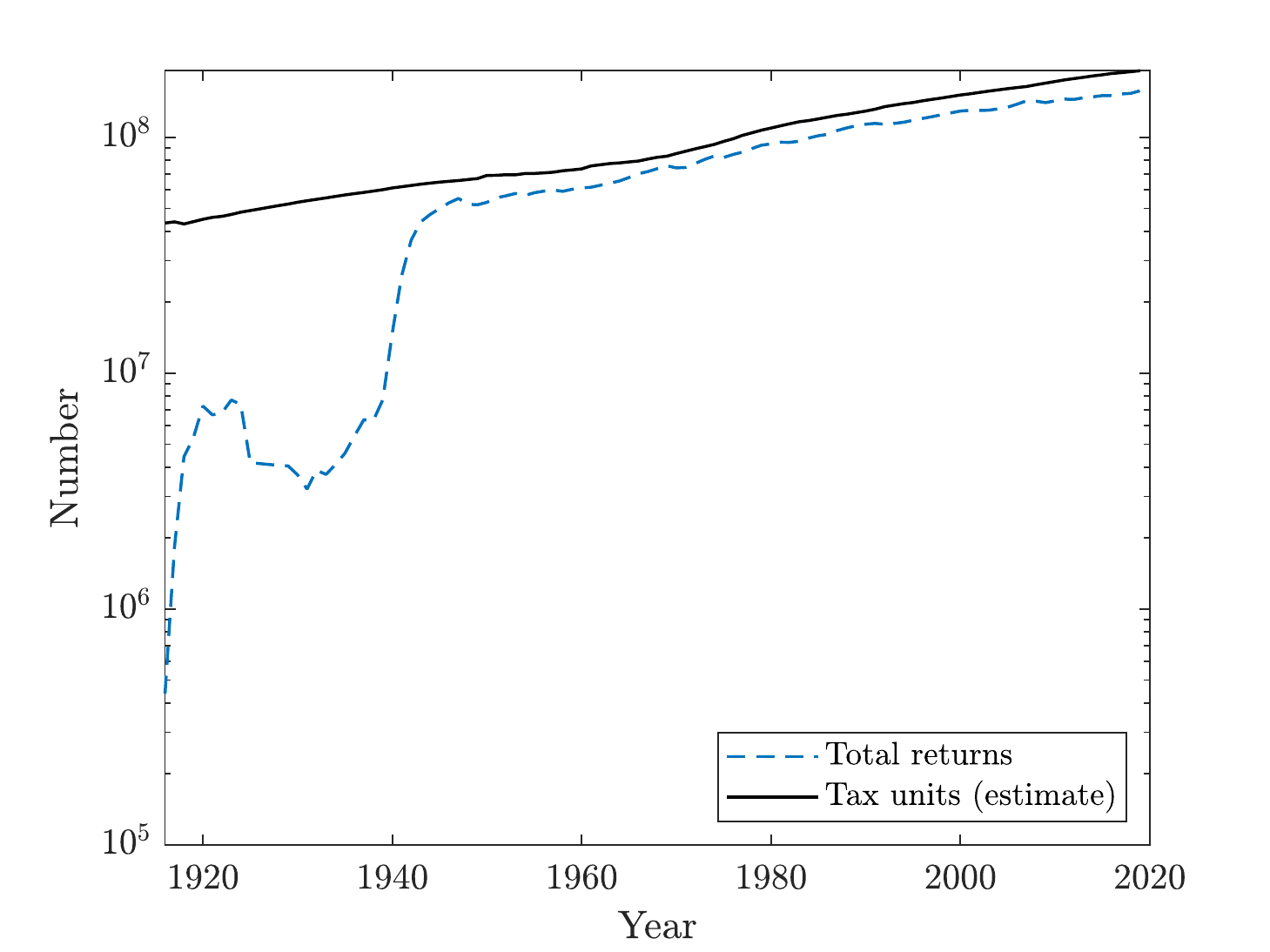}
\caption{Potential tax units.}\label{fig:tax_unit}
\end{subfigure}
\caption{Adult population, total and joint returns, and potential tax units.}
\end{figure}

To estimate the sample size (the number of tax filers if all tax units were forced to file for taxes), we take the following approach. We first collect data on the total number of tax returns ($T$), the total number of tax returns of married couples filing jointly ($J$), the adult population ($A$), and the number of married couples ($M$).\footnote{We obtain the intercensal (July) estimates of the United States resident population by age and sex from the Census Bureau (\url{https://www2.census.gov/programs-surveys/popest/tables/}) and define the adult population as the number of individuals with age 20 or above. We obtain the number of married women with age 15 or above from the Census Bureau (\url{https://www2.census.gov/programs-surveys/demo/tables/families/time-series/marital/ms1.xls}) post-1970 and from \emph{Historical Statistics of the United States, Colonial Times to 1970} (\url{https://www2.census.gov/library/publications/1975/compendia/hist_stats_colonial-1970/hist_stats_colonial-1970p1-chA.pdf}) pre-1970 and define it to be the number of married couples. We use linear interpolation in logarithmic scale to interpolate for intercensal years.} Figure \ref{fig:adult} shows these numbers as well as $T+J$, which would equal $A$ if only adults file for taxes and every adult files for taxes.\footnote{Note that the number of total returns is $T=S+J$, where $S$ is the number of single returns. Thus if every adult files for taxes, it must be $A=S+2J=T+J$.} $T+J$ is about 90\% of $A$ since 1950, suggesting that non-filers consist of only about 10\% of the adult population. Before 1940, $T+J$ is one or two orders of magnitude smaller than $A$.\footnote{The dramatic increase in the number of tax filers during the early 1940s can be likely attributed to the Revenue Act of 1942, whose purpose was to raise revenues to finance World War II efforts by increasing the tax base and tax rates. This amendment reduced the minimum amount of gross income for which a tax return is required (as well as a personal exemption) from \$1,500 to \$1,200 for married couples, increasing the tax base. See p.~3 of \emph{Statistics of Income for 1942} available at \url{https://www.irs.gov/pub/irs-soi/42soireppt1ar.pdf} for the details on the Revenue Act of 1942.}

If non-filers would mostly file individually (not jointly) if they were forced to file, then the number of potential tax units can be estimated as $A-J$, at least after 1950. To test this conjecture, we compute the fraction of joint returns among all returns across the income distribution for 1950, 1970, and 2019, which is shown in Figure \ref{fig:frac_joint_fractile}.  We see that low (high) income earners tend to file separately (jointly). We assume that non-filers do not file for taxes because their incomes are below the exemption level (and would file separately if forced to file), and thus we estimate the sample size as $A-J$ post-1950.

This estimate cannot be used pre-1950 because only a small fraction of adults filed for taxes in that period according to Figure \ref{fig:adult}. To estimate the number of potential joint returns pre-1950, in Figure \ref{fig:frac_MJ} we plot the number of married couples divided by adult population ($M/A$) and the number of joint returns divided by adult population ($J/A$), which seem to have a clear common trend post-1950. We thus use the post-1950 data and regress $\log(J/A)$ on $\log(M/A$) ($R^2=0.989$) and compute the fitted value $\hat{J}$ pre-1950 from the OLS estimates. Finally, we estimate the sample size as $A-\hat{J}$ pre-1950. Figure \ref{fig:tax_unit} shows our final estimate of the sample size. Appendix \ref{subsec:robust_n} considers an alternative definition of the sample size and shows that our results are robust.

\section{Results}\label{sec:results}

\subsection{Tail behavior of income}

Before we estimate the income Pareto exponents, we first check whether the Pareto tail assumption is reasonable. The probability density function of the Pareto distribution (with Pareto exponent $\alpha>1$ and minimum size normalized to 1) is given by $f(y)=\alpha y^{-\alpha-1}$ for $y\ge 1$. The fraction of the population with income of at least $y$ is
\begin{equation*}
p=\int_y^\infty f(y)\diff y=y^{-\alpha}.
\end{equation*}
Eliminating $y$, the total income accruing to the top $p$ fractile of the population is
\begin{equation*}
Y(p)=\int_y^\infty yf(y)\diff y=\frac{\alpha}{\alpha-1}y^{1-\alpha}=\frac{\alpha}{\alpha-1}p^{1-1/\alpha}.
\end{equation*}
Therefore the income share of the top $p$ fractile is
\begin{equation}
S(p)=\frac{Y(p)}{Y(1)}=p^{1-1/\alpha}.\label{eq:Sp}
\end{equation}
The relation \eqref{eq:Sp} shows that top income shares and fractiles are linear in log-log scale with a slope of $1-1/\alpha$. A straightforward calculation reveals that the same relation (but with a different intercept) holds if the income distribution is not exactly Pareto but only has a Pareto upper tail, as long as $p$ is small enough.

\begin{figure}[!htb]
\centering
\includegraphics[width=0.7\linewidth]{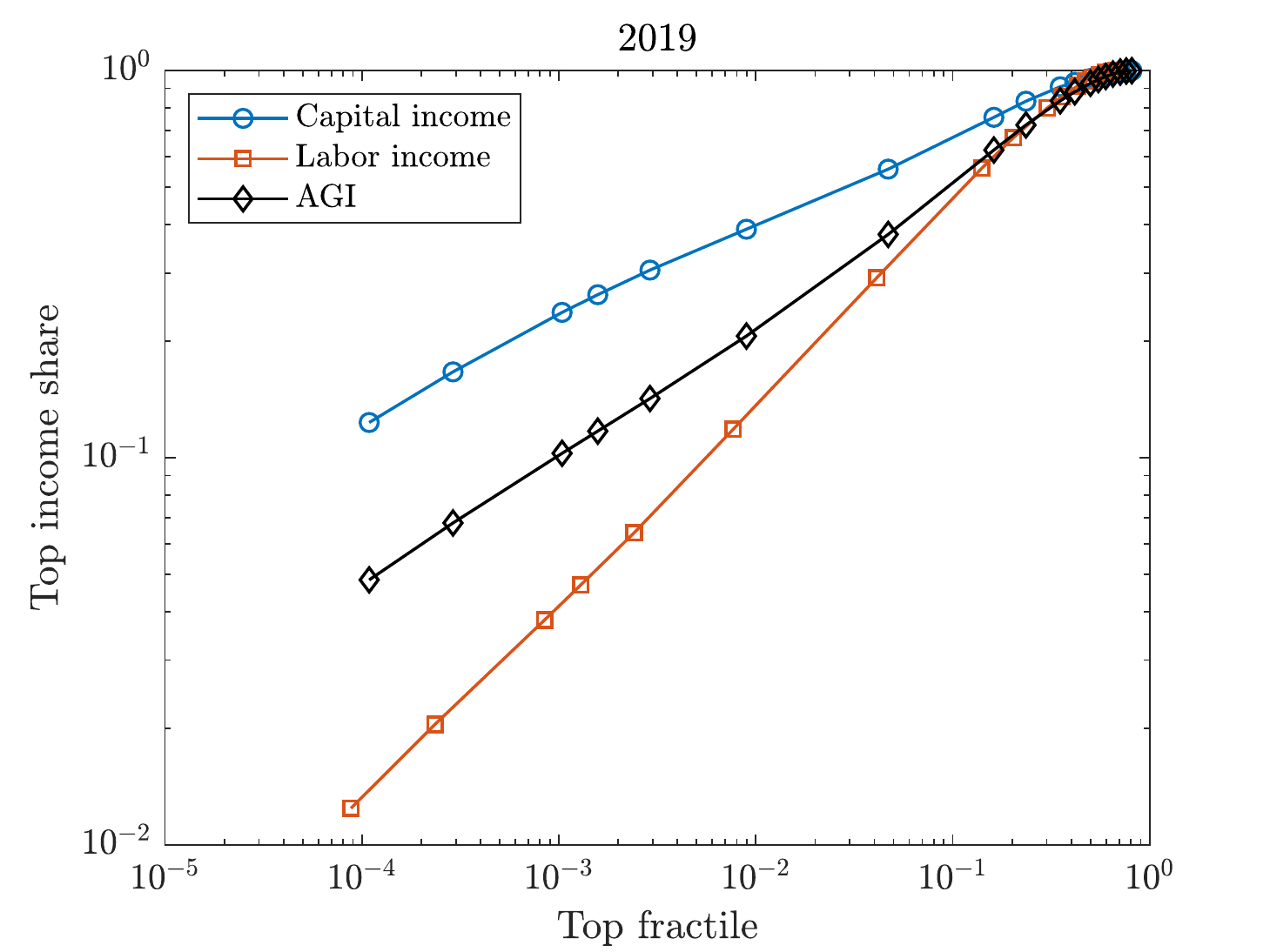}
\caption{Top income shares in 2019.}\label{fig:logshare}
\caption*{\footnotesize Note: The top fractiles are defined by using the potential tax units (adult population minus the number of joint returns) as the denominator. The top income shares are defined by using the sum across all returns with positive AGI as the denominator.}
\end{figure}

Figure \ref{fig:logshare} plots the top shares (among all tax filers with positive AGI) of AGI, capital income, and labor income in 2019. Consistent with a Pareto upper tail and \eqref{eq:Sp}, the graphs show straight-line patterns for income above the top 1--10\%. The fact that the slope for labor income is steeper than that of capital income suggests that labor income has a larger Pareto exponent. All other years exhibit similar patterns.

\subsection{Capital income Pareto exponents}\label{subsec:capital_exponent}

We estimate Pareto exponents of AGI and capital income using the minimum distance method of \cite{TodaWang2021JAE} detailed in Appendix \ref{sec:estim}.

As Pareto's law applies only to the upper tail, we need to take a stance on the income groups used in the estimation (the number $L+1$ in Appendix \ref{sec:estim}). The literature typically uses the largest 5\% observations for estimating the Pareto exponent.\footnote{See the discussions around Footnotes 8 and 9 in \cite{deVriesTodaLIS}.} In general, choosing a smaller threshold reduces the sampling error (due to the larger sample size used for estimation) but increases the bias due to potential model misspecification. Because our sample sizes are large and sampling error is not an issue, to be conservative we use the largest 1\% observations (choose the largest $L$ such that the top $L+1$ income groups are within the top 1\%; see Appendix \ref{subsec:robust_L} for robustness). Figure \ref{fig:ngroup} shows the number of income groups in the original data set ($K$) and those used for estimation ($L+1$). Before 1952, the tabulated summaries are relatively detailed and $K$ ranges between 30 and 50. Since 1952, the information is more compressed and $K$ tends to be between 15 and 30.

\begin{figure}[!htb]
\centering
\includegraphics[width=0.7\linewidth]{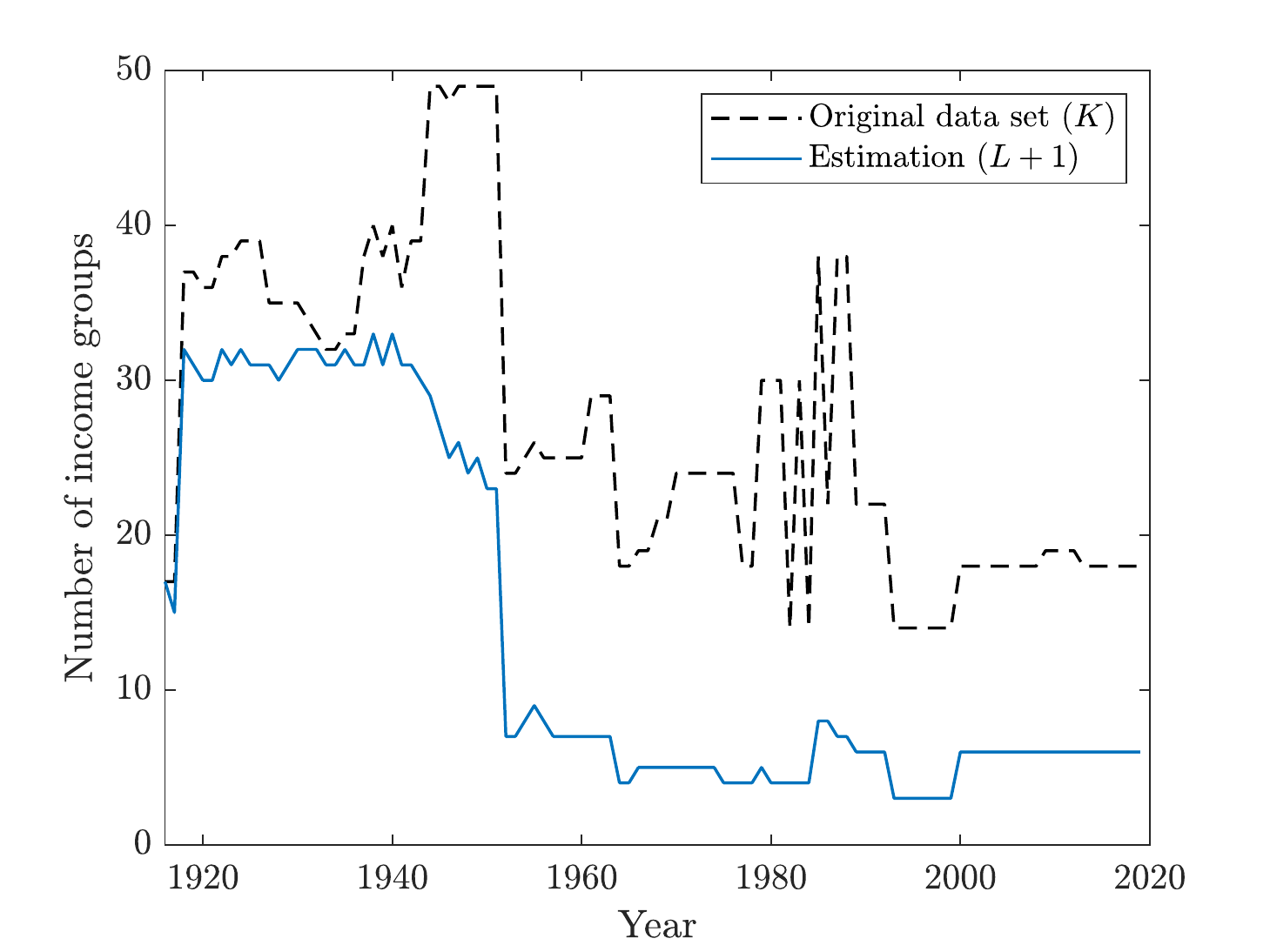}
\caption{Number of income groups.}\label{fig:ngroup}
\end{figure}

For estimating the capital income Pareto exponents, we need one additional assumption. Note that income groups in Table \ref{t:IRS2019} are defined by the rank in AGI. However, because for high income earners AGI and capital income are nearly identical (Figure \ref{fig:average_income}), we assume that tax units remain in the same income group defined by AGI even after rearranging with respect to capital income. For example, if tax unit $i$ belongs to group $k$ and tax unit $i'$ belongs to group $k'>k$, so $Y_i>Y_{i'}$, then we also have $Y_i^c>Y_{i'}^c$, where $Y_i^c$ denotes the capital income of tax unit $i$. Thus, although we do not require that the capital income is ranked in the same way as AGI \emph{within} the same income group $k$, we assume the same ranking \emph{across} different income groups $k'>k$.

\begin{figure}[!htb]
\centering
\includegraphics[width=0.7\linewidth]{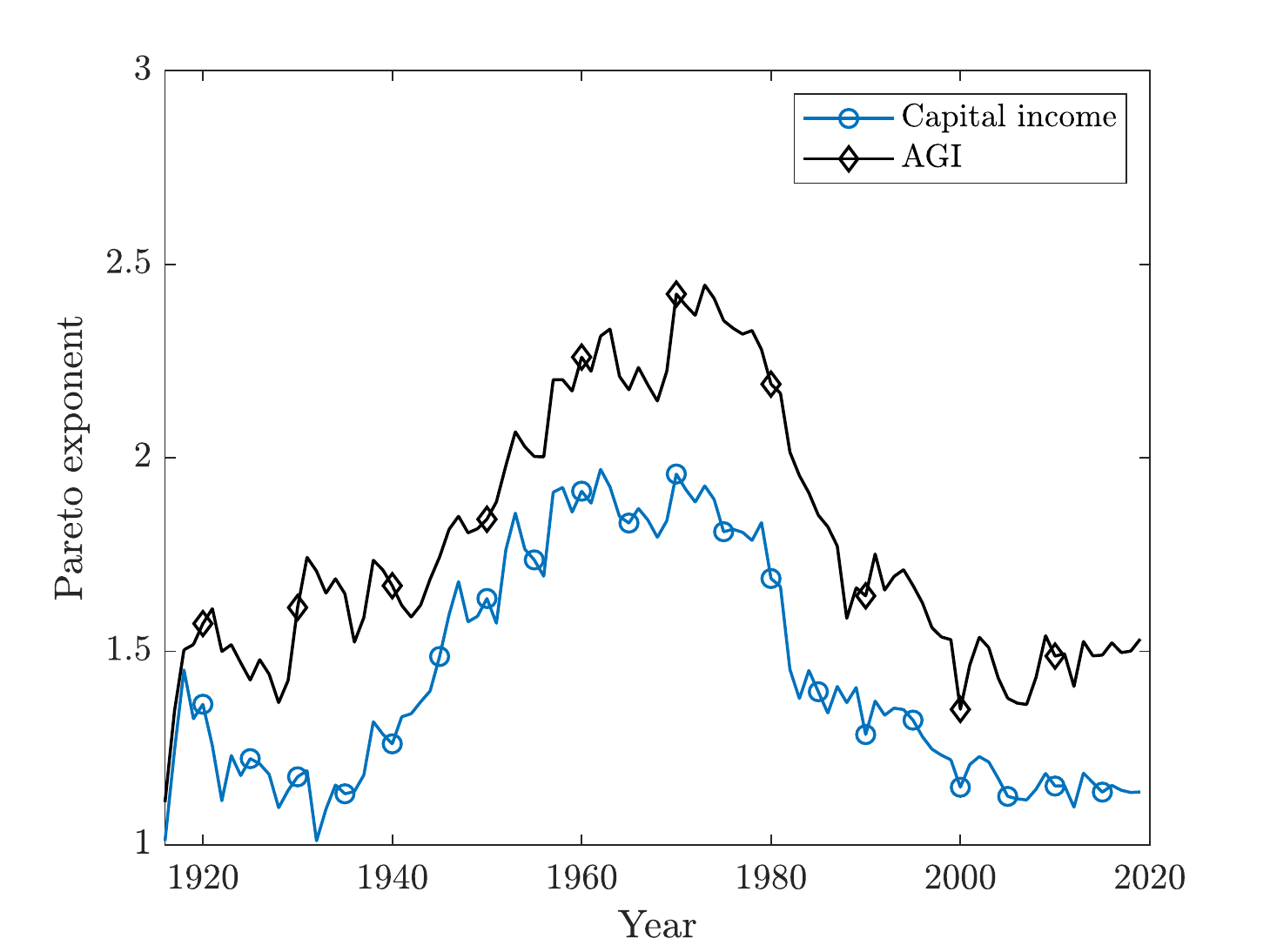}
\caption{Income Pareto exponents in United States.}\label{fig:alpha}
\end{figure}

Figure \ref{fig:alpha} plots the estimated Pareto exponents for AGI and capital income. We report the standard errors in Appendix \ref{subsec:robust_se}, which have the order of magnitude $10^{-3}$.\footnote{The order of magnitude $10^{-3}$ can actually be inferred without doing any computation. The number of tax units is about one hundred million ($10^8$). Since we use the top 1\% and the standard error inversely scales with the square root of the sample size used in the estimation, the order of magnitude of the standard error is $(10^8\times 0.01)^{-1/2}=10^{-3}$.} Figure \ref{fig:alpha} suggests that the sample period can be divided into three sub periods, pre-1940, 1940--1985, and post-1985. The capital income Pareto exponent is roughly constant at 1.2 pre-1940 and post-1985, and shows an inverse U-shaped pattern during 1940--1985, plateauing at 2.0 in 1960--1970.

\begin{figure}[!htb]
\centering
\includegraphics[width=0.7\linewidth]{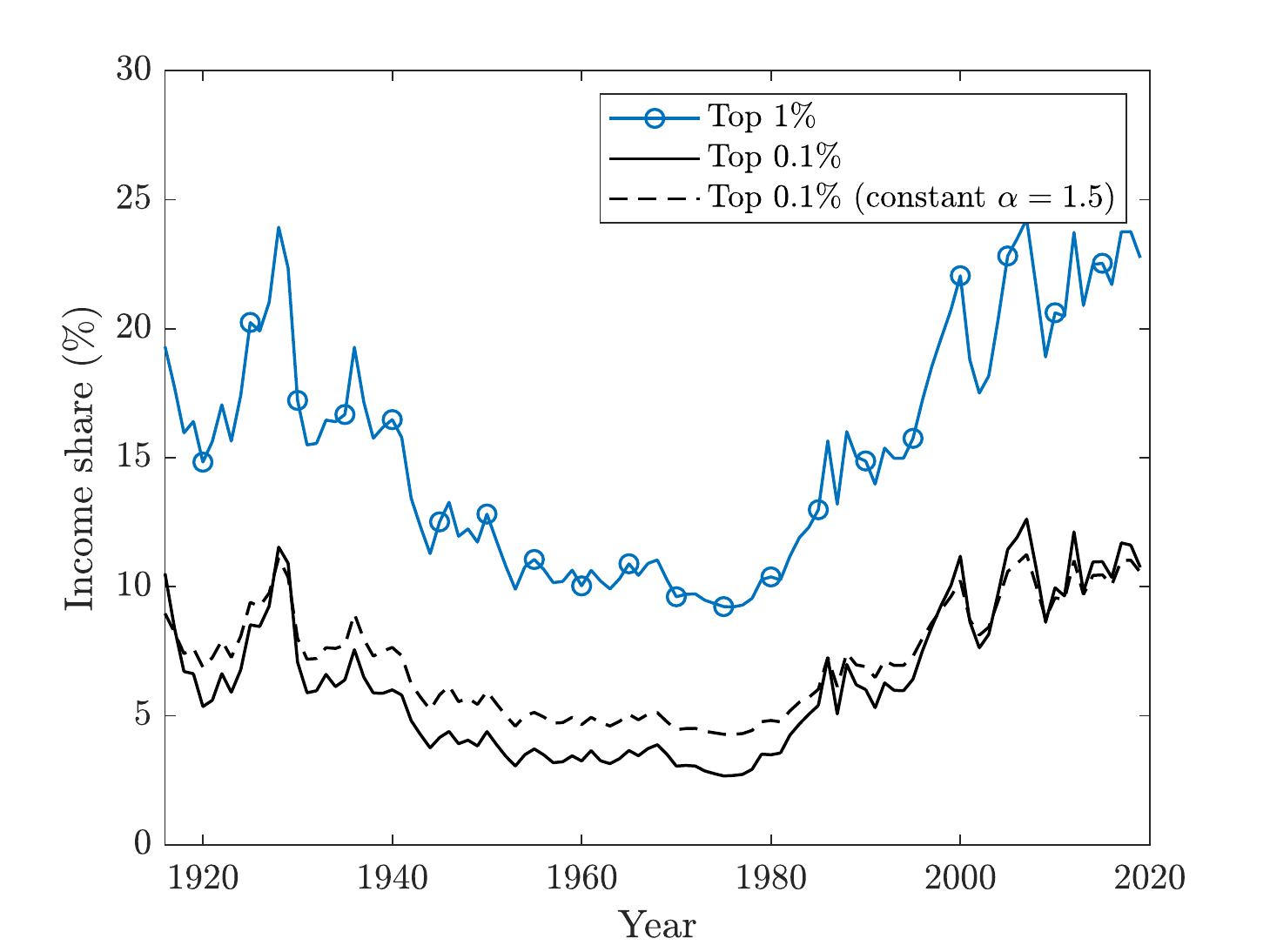}
\caption{Top income shares in United States.}\label{fig:top1share}
\end{figure}

What is remarkable here is that none of the Pareto exponents show a clear trend post-1985. Compare this to the top 1\% and 0.1\% income shares constructed by \cite{PikettySaez2003}, shown in Figure \ref{fig:top1share}.\footnote{\label{fn:PSZ}The updated top income shares of \cite{PikettySaez2003} are from Tab TD1 of the spreadsheet \url{https://gabriel-zucman.eu/files/PSZ2022AppendixTablesII(Distrib).xlsx}.} Here the top 1\% income share has almost doubled from 13.0\% in 1985 to 22.8\% in 2019. This finding suggests that the rise in income inequality during the past three or four decades concerns the inequality between the rich and the rest, and not the top tail inequality. To test this conjecture, the dashed line in Figure \ref{fig:top1share} shows the top 0.1\% share implied by the top 1\% share and an AGI Pareto exponent equal to $\alpha=1.5$, which is the post-1985 average in Figure \ref{fig:alpha}.\footnote{To construct the implied top 0.1\% income shares, we note that when the income distribution has a Pareto upper tail with exponent $\alpha$, then the top $p$ fractile income share takes the form $S(p)=Ap^{1-1/\alpha}$ for some constant $A>0$. Therefore if $0<p<q\ll 1$, we have the relation $S(p)=(p/q)^{1-1/\alpha}S(q)$, which we apply for $(p,q)=(0.001,0.01)$.} The fact that the implied top 0.1\% shares are similar to the actual top 0.1\% shares since 1985 suggests that top tail income inequality has been stable during this period.

\subsection{Labor income Pareto exponents}\label{subsec:labor_exponent}

We next estimate the labor income Pareto exponents. There are several issues with doing so. When we estimated the capital income Pareto exponents in Section \ref{subsec:capital_exponent}, we assumed that capital income is ordered in the same way as AGI across income groups, at least for top income earners. Although this assumption is reasonable for capital income because capital income is nearly identical to AGI for high income earners (Figure \ref{fig:average_income}), it is suspect for labor income because the fraction of labor income is small for high income earners. Second, although SOI publishes the size distribution of salaries and wages (labor income) for some years, this information is limited to the period 1927--1978.

\begin{figure}[!htb]
\centering
\begin{subfigure}{0.48\linewidth}
\includegraphics[width=\linewidth]{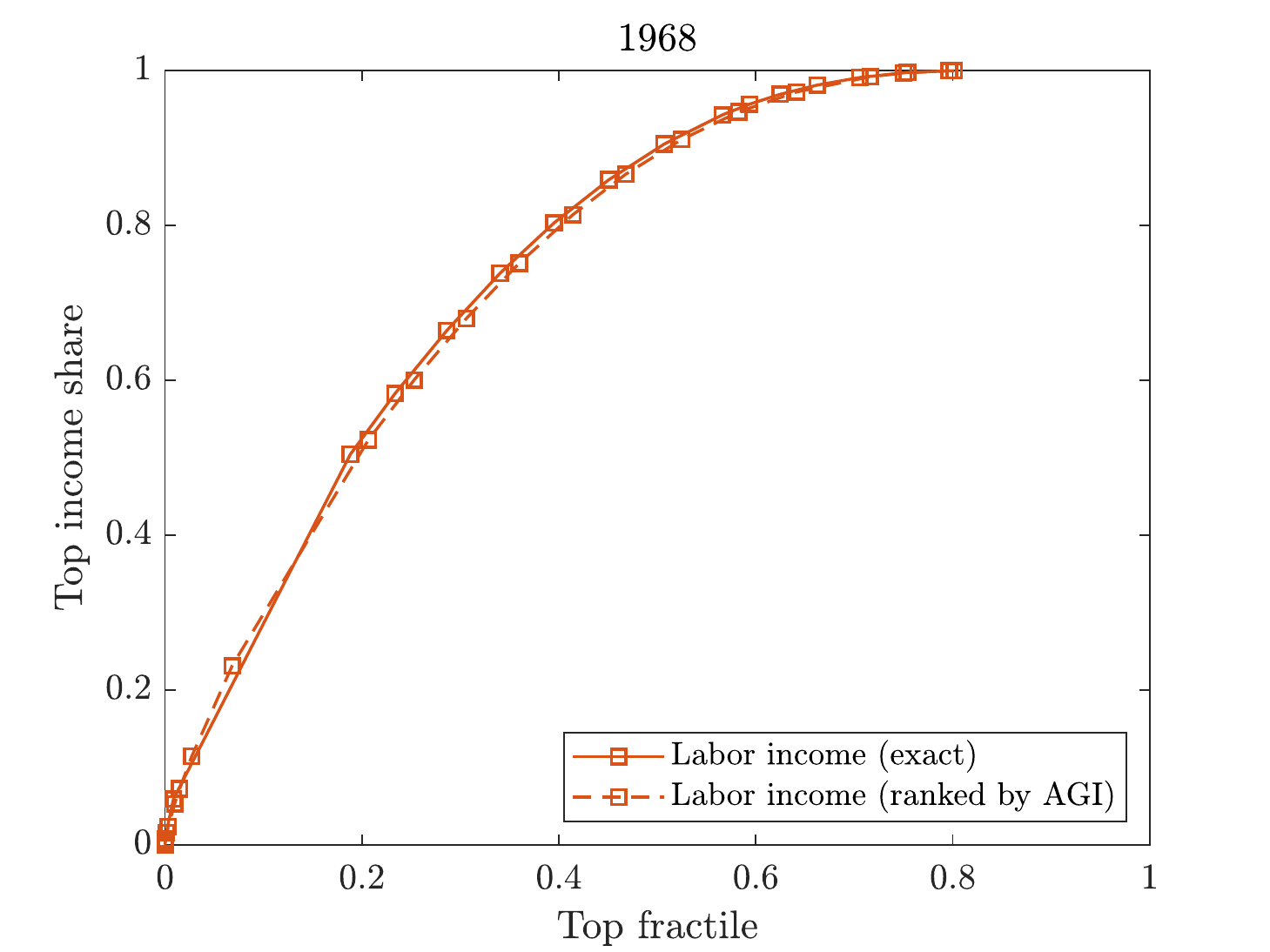}
\caption{Original scale.}\label{fig:log_share_wage_original}
\end{subfigure}
\begin{subfigure}{0.48\linewidth}
\includegraphics[width=\linewidth]{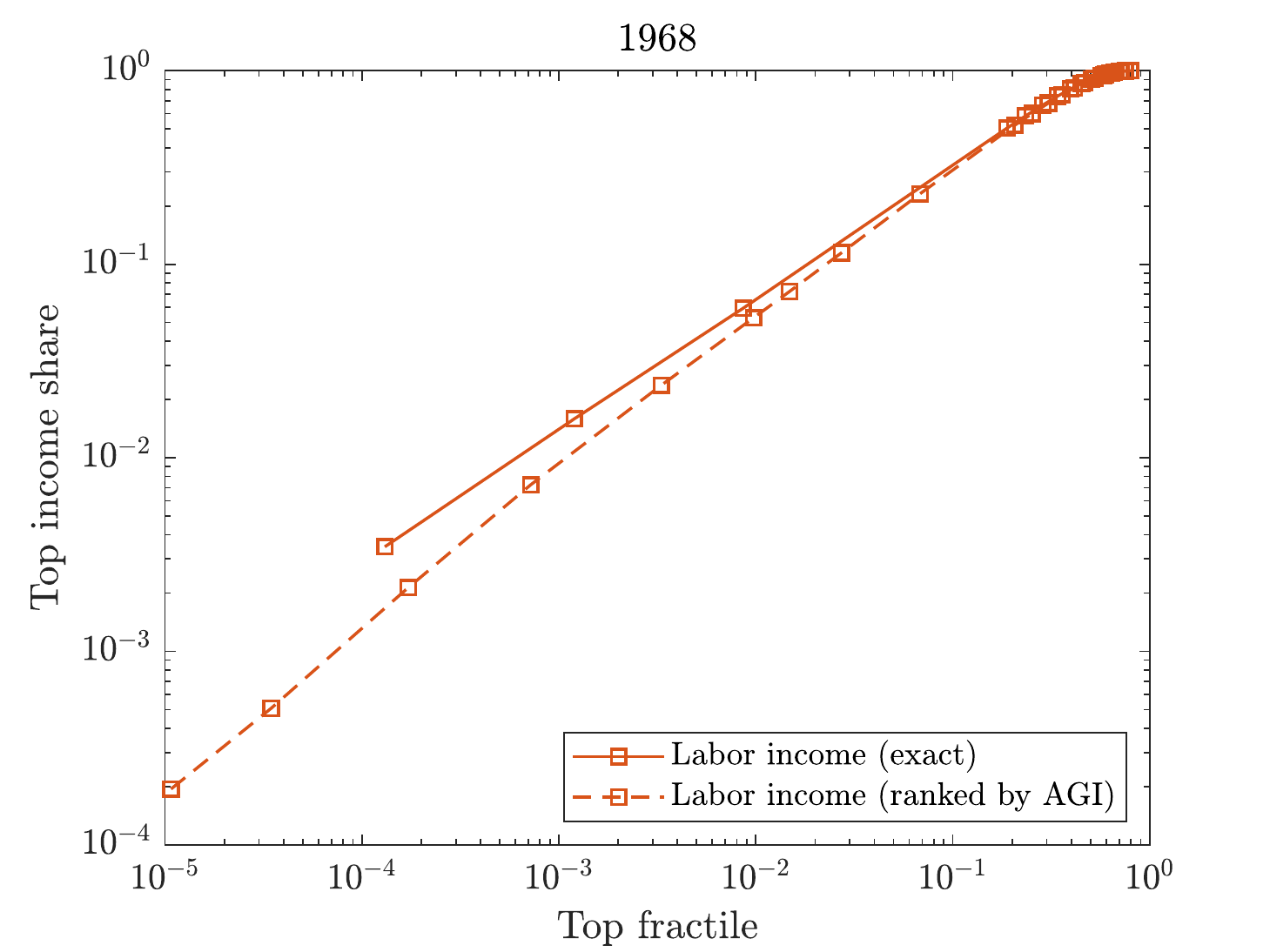}
\caption{Log-log scale.}\label{fig:log_share_wage_loglog}
\end{subfigure}
\caption{Top labor income shares in 1968.}\label{fig:logshare_wage}
\caption*{\footnotesize Note: Top labor incomes are ranked either exactly (solid line) or by AGI (dashed line). The left (right) panel plots the top income shares in the original (log-log) scale.}
\end{figure}

To address these concerns, Figure \ref{fig:logshare_wage} plots the top labor income shares in 1968, both when top labor incomes are exactly ranked (solid line) and ranked by the size of AGI (dashed line). We choose the year 1968 because a detailed table on the joint size distribution of AGI and labor income as well as tables on cumulative sums of top incomes (both AGI and labor income) are available.\footnote{See Table 1.9 on p.~27 of \emph{Statistics of Income for 1968} available at \url{https://www.irs.gov/pub/irs-soi/68inar.pdf}, which reports the number of tax returns grouped by the size of adjusted gross income and the size of salaries and wages. The original table combines cells when the number of returns is small (\eg, low AGI but high labor income, which is uncommon). In those cases we assign the number to the highest AGI group among the combined groups and assign zero to other groups.} According to Figure \ref{fig:log_share_wage_original} (original scale), the top labor income shares look similar in the body of the size distribution regardless of how we define the ranking. However, when we plot the top labor income shares in a log-log scale (Figure \ref{fig:log_share_wage_loglog}), ranking salaries and wages using AGI makes the slope steeper, suggesting that high AGI earners tend to have relatively small labor income.

\begin{figure}[!htb]
\centering
\begin{subfigure}{0.48\linewidth}
\includegraphics[width=\linewidth]{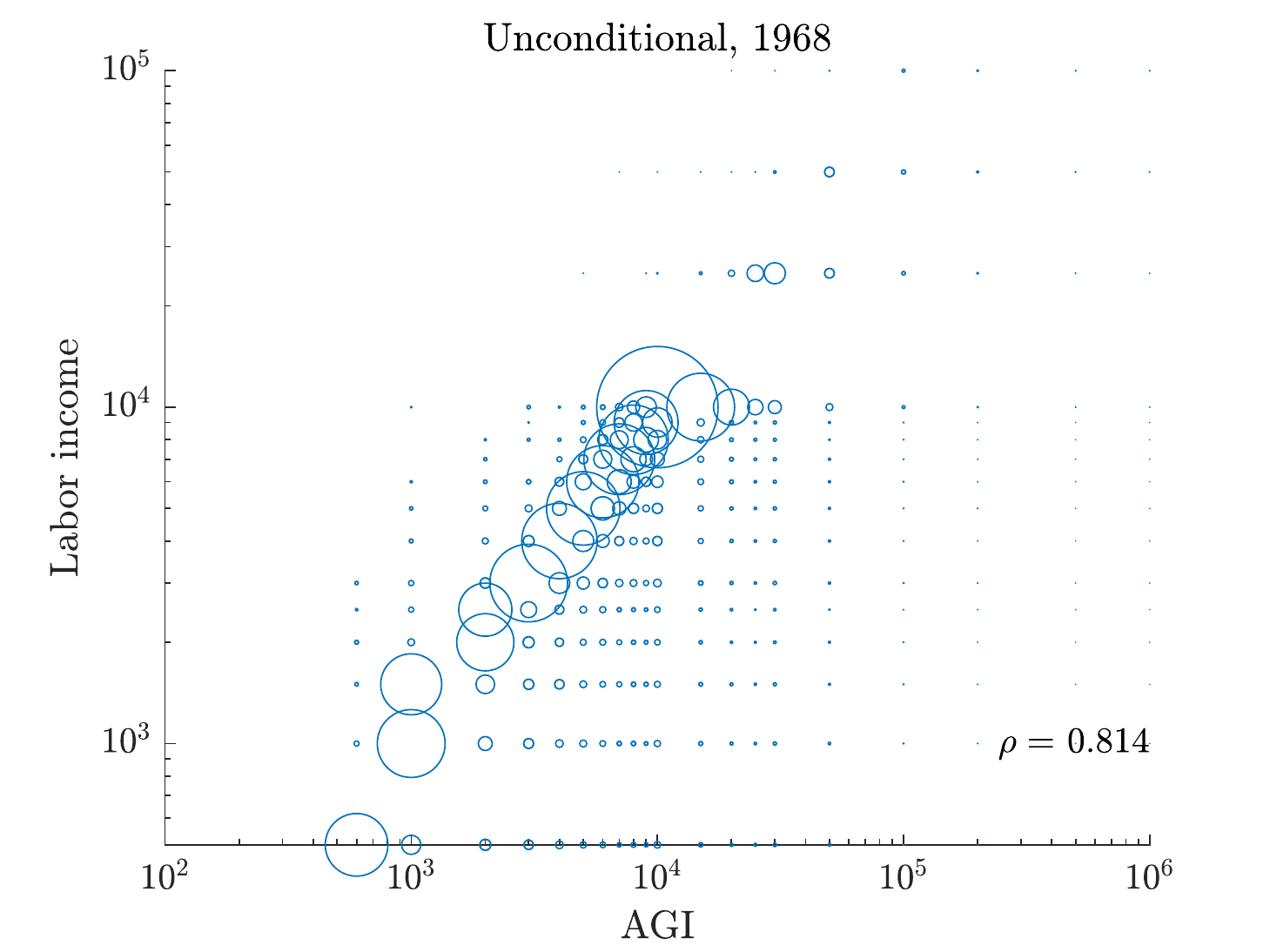}
\caption{Unconditional.}\label{fig:scatter_wage_unc}
\end{subfigure}
\begin{subfigure}{0.48\linewidth}
\includegraphics[width=\linewidth]{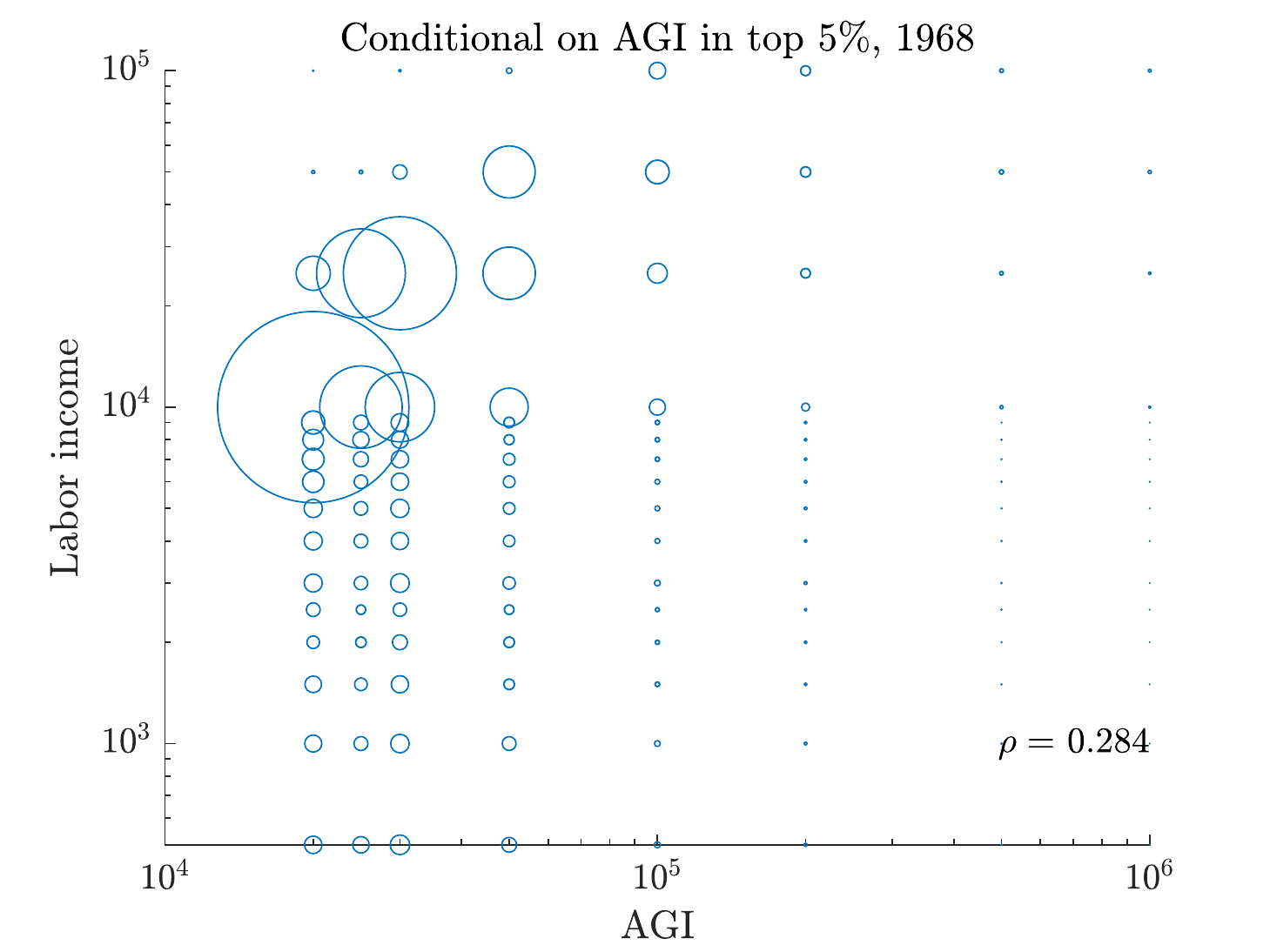}
\caption{Conditional on AGI in top 5\%.}\label{fig:scatter_wage_cond}
\end{subfigure}
\caption{Joint distribution of AGI and labor income, 1968.}\label{fig:scatter_wage}
\caption*{\footnotesize Note: The marker sizes are proportional to the frequency.}
\end{figure}

To investigate this issue further, Figure \ref{fig:scatter_wage} shows the joint distribution of AGI and labor income on a log-log scale. The horizontal and vertical axes show the lower thresholds of AGI and salaries and wages and the marker sizes are proportional to the frequency. According to Figure \ref{fig:scatter_wage_unc}, which shows the entire (unconditional) joint distribution, AGI and labor income are highly correlated ($\rho=0.814$). However, when we plot the joint distribution conditional on high AGI (in the top 5\%) in Figure \ref{fig:scatter_wage_cond}, the correlation is much weaker ($\rho=0.284$).

\begin{figure}[!htb]
\centering
\includegraphics[width=0.7\linewidth]{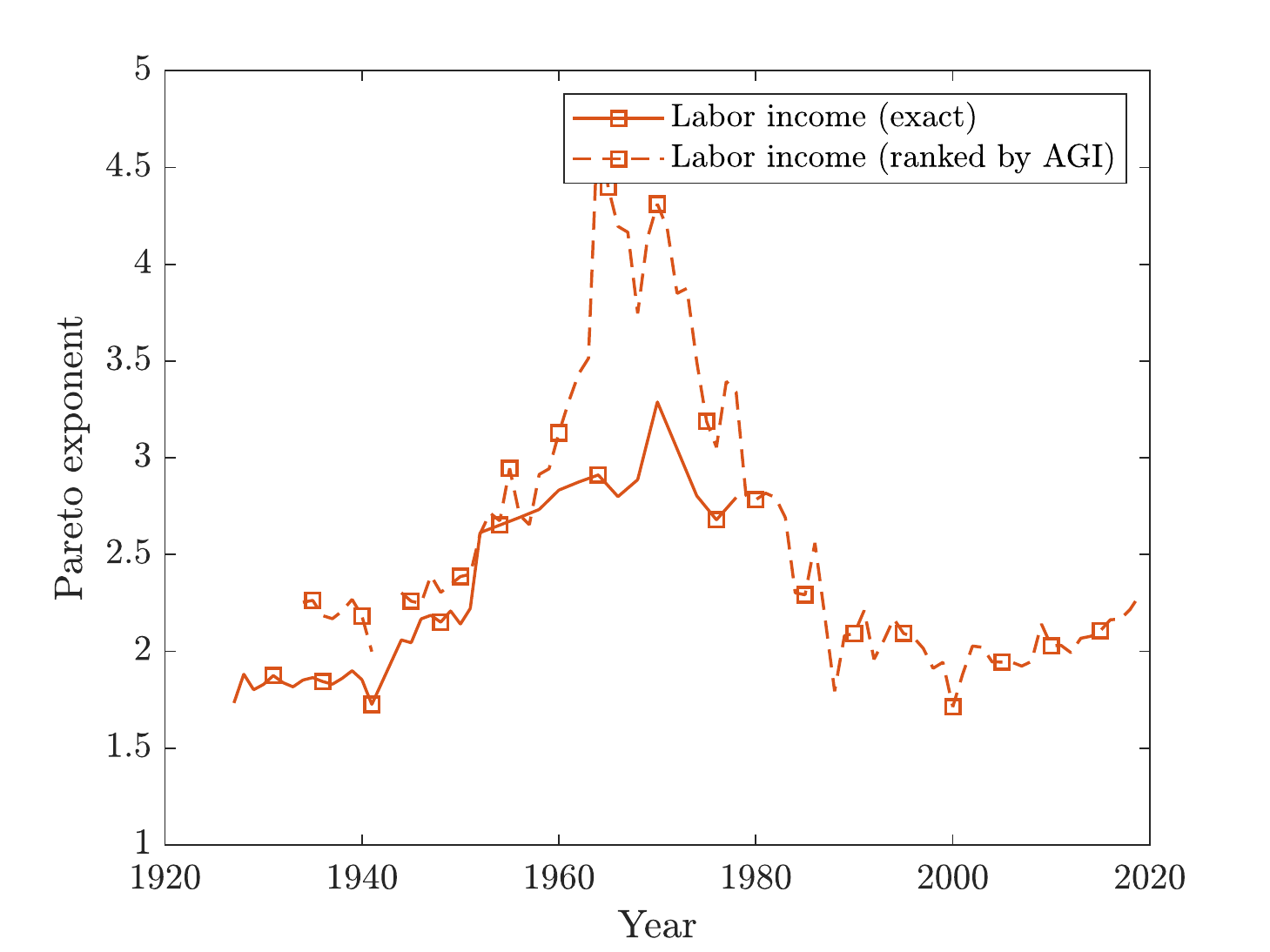}
\caption{Labor income Pareto exponents.}\label{fig:alpha_labor}
\end{figure}

Our analysis from Figures \ref{fig:logshare_wage} and \ref{fig:scatter_wage} suggests that the assumption that labor income and AGI are ordered in the same way across income groups roughly holds for low to middle income earners but does not hold for high income earners. However, it is for high income earners that we need this assumption to estimate labor income Pareto exponents using the method of \cite{TodaWang2021JAE}. Given this limitation, in Figure \ref{fig:alpha_labor} we plot two estimates of the labor income Pareto exponent. The first (solid line) is the maximum likelihood (ML) estimate (see Appendix \ref{subsec:estim_ML} for details) using the exact size distribution of labor income whenever available. Although these data require no assumption for estimation (except that the distribution has a Pareto upper tail and the observations are \iid), they are available only for a subset of years during the period 1927--1978. The second estimate (dashed line) is the \cite{TodaWang2021JAE}  (TW) estimate using the top labor income ranked by AGI, ignoring the relatively low correlation of AGI and labor income for high AGI earners. These data are available since 1934 excluding 1942 and 1943 only but the assumption is suspect.

According to Figure \ref{fig:alpha_labor}, the ML and TW estimates are qualitatively similar in that they are low and stable before 1940, start rising around 1950, and peak around 1970. However, the ML estimates are uniformly smaller than the TW estimates, consistent with Figure \ref{fig:log_share_wage_loglog}. This suggests that the labor income Pareto exponents post-1980 (for which ML is inapplicable due to the nonexistence of data) are likely slightly smaller than the TW estimates in Figure \ref{fig:alpha_labor}.

\subsection{Comparison to existing estimates}

How do our Pareto exponent estimates compare to those reported in the literature? Although it is well known that the income distribution has a Pareto upper tail, it is not easy to find empirical estimates because few papers distinguish capital and labor income. \citet[Table A-1]{FeenbergPoterba1993} and \citet[Table 13A.23]{AtkinsonPiketty2010} only report the Pareto exponents for adjusted gross income (AGI), which are reviewed in Appendix \ref{subsec:robust_estim}. The empirical application of \cite{TodaWang2021JAE} estimates the Pareto exponents over the past century, but they use as input the top income shares constructed by \cite{PikettySaez2003}, which is not raw data, and the analysis is limited to AGI due to data availability.

\begin{figure}[!htb]
\centering
\includegraphics[width=0.7\linewidth]{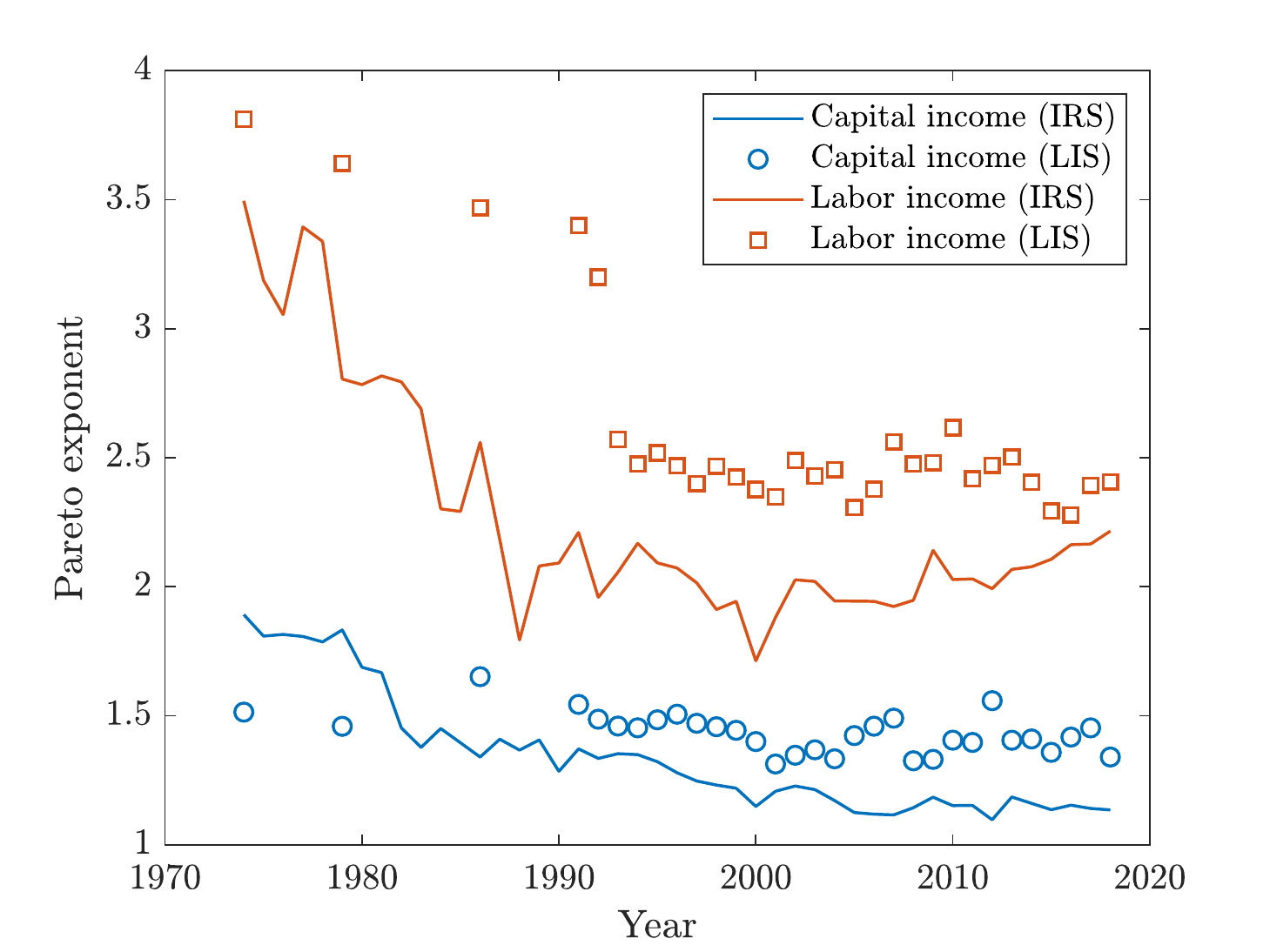}
\caption{Pareto exponents estimated from administrative and survey data.}\label{fig:alpha_LIS}
\caption*{\footnotesize Note: The solid lines show the estimates from IRS (administrative) data taken from Figures \ref{fig:alpha} and \ref{fig:alpha_labor}. The markers show the estimates from LIS (survey) data taken from Table 2 of the Online Appendix of \cite{deVriesTodaLIS}.}
\end{figure}

\citet[Figure 2]{Toda2012JEBO} estimates the U.S. labor income Pareto exponents for males using the \emph{Panel Study of Income Dynamics (PSID)} for 1968--1993 and the \emph{Current Population Survey (CPS)} for 2000--2009. He finds that the labor income Pareto exponent declined from about 4 to 3 over the period 1968--1993 and has been stable at around 2.5 during 2000--2009. These values are qualitatively similar but larger than those in Figure \ref{fig:alpha_labor}. Table 2 in the Online Appendix of \cite{deVriesTodaLIS} reports capital and labor income Pareto exponents for U.S. households for the years 1974, 1979, 1986, and 1991--2018 estimated from the \emph{Luxembourg Income Study} database, which is based on the \emph{Current Population Survey}. Figure \ref{fig:alpha_LIS} shows the capital and labor income Pareto exponents estimated from the IRS (administrative tax record) and LIS (survey) data. Although both estimates have roughly the same order of magnitude, the LIS estimates are biased upwards (the tails of the income distribution appear to be thinner).

This finding suggests that high income earners tend to either have a lower response rate or underreport incomes in surveys. A key advantage of our empirical approach is that we only use tabulated summaries of incomes from tax authorities, which are likely more accurate than survey data and publicly available for a long sample period. The fact that our newer (and presumably more accurate) estimates are smaller than those in the literature suggests that the top tail capital and labor income inequality is higher than previously thought.

\section{Concluding remarks}

In this paper, using the best publicly available data (tabulated summaries of incomes from IRS that contain over one hundred million observations) and the best econometric method (that requires only the observation of the number of tax filers in income groups and their group averages), we estimated income Pareto exponents in the U.S. for 1916--2019. To our knowledge our paper is the first to systematically report capital and labor income Pareto exponents separately for a long sample period. Our new estimates of 1.2 for capital income and 2 for labor income are smaller (income inequality higher) than previous estimates and are stable post-1985. The new capital income Pareto exponent estimate of 1.2 suggests that the impact of wealthy agents on the aggregate economy is twice as large as implied by previous estimates. A policy implication of our empirical findings is that, because high incomes are mostly capital income, taxation of capital income is likely more effective for redistribution purposes.


\newpage

\appendix

\section{Details on data}\label{sec:detail}

Our primary data source is the \emph{Statistics of Income (SOI) Individual Income Tax Returns Publication 1304} from the United States Internal Revenue Service. The data since 1993 are available as Excel spreadsheets, while  the data before 1993 are only available as PDFs of scanned copies of SOI.\footnote{See Table 1.4 at \url{https://www.irs.gov/statistics/soi-tax-stats-individual-income-tax-returns-complete-report-publication-1304-basic-tables-part-1} for data since 1993 and \url{https://www.irs.gov/statistics/soi-tax-stats-archive} for data before 1993.}

For years since 1993, we observe the information in a format similar to Table \ref{t:IRS2019}, although the number of income groups and thresholds for adjusted gross income (AGI) differ across years (see Figure \ref{fig:ngroup}). For years before 1993, the format is less consistent and we do our best to record the data as consistently as possible using the following procedures.
\begin{itemize}
\item For years 1985--1992, SOI includes two tables of the same variables, one in Section 1 whose AGI thresholds are finer for high income and another in Section 2 (Table 1.4) whose thresholds are finer for low income. Whenever possible we combine the two tables to use the finest thresholds.
\item For years 1940--1960 and 1966--1969, SOI does not report the number of returns and total income for salaries and wages (Table \ref{t:IRS2019}, Columns (4) and (5)) among \emph{all} returns, but only for \emph{taxable} and \emph{nontaxable} returns separately. In those cases we impute the number of returns and total income among all returns by adding taxable and nontaxable returns whenever possible. The only income group we cannot apply this procedure is the highest nontaxable income group, for which all income groups above the lower AGI threshold are lumped together. However, the number of returns and total income of this group is quite small so we do not expect material impact on the results.
\item For years before 1952, the AGI thresholds tend to be fine (see Figure \ref{fig:ngroup}) and the number of returns (Table \ref{t:IRS2019}, Columns (2) or (4)) is sometimes zero for a particular income group. In those cases, we combine it with the next nonzero group to avoid zero observations, which is necessary for applying the statistical method.
\item For years 1941--1943, we combine income groups with AGI less than \$3,000 together with tax filers who used Form 1040A because the Form 1040A lacks the detailed breakdown of income (except that AGI must be less than \$3,000).
\item For years 1916, 1917, and 1927--1936, the AGI thresholds (Table \ref{t:IRS2019}, Column (1)) used to report AGI and salaries and wages differ. In those cases, we combine income groups so that we use the finest common thresholds. (For years 1927--1936, this procedure applies only to the bottom five income groups. However, for 1916 and 1917, the AGI thresholds used for salaries and wages are significantly more coarse.)
\item For years 1934--1936, SOI does not report the number of returns for salaries and wages with AGI less than \$5,000. For those cases we use the number of returns for AGI instead.
\item For years 1916--1933, 1942, and 1943, SOI does not report the number of returns for salaries and wages (Table \ref{t:IRS2019}, Column (4)) at all. For those cases we can only estimate the Pareto exponents of AGI and capital income.
\end{itemize}

For years before 1993, because we manually entered the numbers into spreadsheets, human errors are inevitable. (For each year, we need to enter about 100 numbers with about 10 digits.) To ensure the accuracy of the numbers, we compared the total number computed from the spreadsheet and that copied from the original SOI tables (the bottom row of Table \ref{t:IRS2019}). When the two numbers were identical within reasonable rounding error (\eg, difference less than 5 in the last digit), we concluded that the data were accurate.

\section{Econometric theory}\label{sec:estim}

Our econometric theory is based on the following assumptions.

\begin{asmp}\label{asmp:iid}
$Y_i$ is \iid with some distribution $F$. 
\end{asmp}
\begin{asmp}\label{asmp:F} 
$F$ has a Pareto upper tail: for some constants $c>0$ and $\alpha>0$, we have
\begin{equation*}
\frac{1-F(y)}{1-F(c)}=(y/c)^{-\alpha}
\end{equation*}
for $y \ge c$.
\end{asmp}

Assumption \ref{asmp:iid} requires that the data set are collected as a simple random sample from the population, which is reasonably satisfied in our case. Assumption \ref{asmp:F} requires that the right tail of the underlying distribution $F$ is Pareto beyond some tail cutoff $c$, which can be considered as a tuning parameter for estimation. The validity of this Pareto tail condition can be justified from both empirical and theoretical perspectives. First, as shown in Figure \ref{fig:logshare}, the top income shares can be well fitted by a linear function of top fractiles, implying that the Pareto tail condition holds in the data set. The simulation studies in \cite{TodaWang2021JAE} suggest that the tail cutoff $c$ being the top 1\% percentile leads to a very good Pareto tail fit. Second, it has been well established in the statistics literature \citep[e.g.,][]{Pickands1975, Smith1987} that all heavy-tailed distributions can be approximated by the Pareto distribution in the right tail as long as they support the extreme value theory. Commonly used continuous distributions are all covered, including Student-t, F, Cauchy, etc.

\subsection{Income thresholds available}\label{subsec:estim_ML}

When the income thresholds $\set{t_k}_{k=1}^K$ that define the income groups and the cumulative number of top income earners $\set{n_k}_{k=1}^K$ are available (which is always the case for AGI as we see in Columns (1) and (2) in Table \ref{t:IRS2019}, sometimes the case for labor income, and never the case for capital income), we can estimate the Pareto exponent $\alpha$ by maximum likelihood.

Let the income thresholds ordered from top to bottom as $\infty=t_0>t_1>\dots>t_K>0$. Let $2\le L\le K$ be the number of income groups used for estimation, assumed to satisfy $t_L\ge c$ in Assumption \ref{asmp:F}. Then the conditional tail probability is given by
\begin{equation*}
\Pr(Y\ge y \mid Y\ge t_L)=(y/t_L)^{-\alpha}.
\end{equation*}
Thus letting $m_k=n_k-n_{k-1}$ the number of tax filers in group $k$ and $P_k(\alpha)=(t_k/t_L)^{-\alpha}-(t_{k-1}/t_L)^{-\alpha}$ be the probability that a tax filer belongs to group $k$ conditional on $Y\ge t_L$, the vector of counts $(m_k)_{k=1}^L$ has a multinomial distribution with probability vector $(P_k(\alpha))_{k=1}^L$. Letting $q_k=m_k/n_L=(n_k-n_{k-1})/n_L$ be the relative frequency of income group $k$ and ignoring terms that do not depend on $\alpha$, the log likelihood function (normalized by the sample size used in the estimation, $n_L$) becomes
\begin{equation}
\ell(\alpha)=\sum_{k=1}^L q_k\log\left((t_k/t_L)^{-\alpha}-(t_{k-1}/t_L)^{-\alpha})\right).\label{eq:logL}
\end{equation}
We can obtain the maximum likelihood estimator $\hat{\alpha}$ by maximizing \eqref{eq:logL}. We omit the discussion of asymptotic theory because it is standard.

\subsection{Income thresholds unavailable}\label{subsec:estim_TW}

When the income thresholds $\set{t_k}_{k=1}^K$ that define the income groups are unavailable (which is the case for capital income) but the cumulative number of top income earners $\set{n_k}_{k=1}^K$ and the partial sums of top incomes $\set{S_{n_k}}_{k=1}^K$ are available, we can estimate the Pareto exponent $\alpha$ using the minimum distance approach of \cite{TodaWang2021JAE}, described below.

Assume $\alpha>1$ and let $\xi=1/\alpha\in (0,1)$ be the tail index. Suppose $K\ge 3$ and let $L$ be a natural number such that $3\le L+1\le K$ and $t_{L+1}\ge c$ in Assumption \ref{asmp:F}. We estimate the Pareto exponent $\alpha$ using only income groups $k=2,\dots,L+1$. (We exclude group $k=1$ because the asymptotic theory is nonstandard when $\alpha\le 2$.) Let $n$ be the sample size. We define the following quantities:
\begin{subequations}\label{eq:defn}
\begin{align}
p_k&\coloneqq n_k/n \in (0,1),\label{eq:pk} \\
\mu(p,q)&\coloneqq \frac{q^{1-\xi}-p^{1-\xi}}{1-\xi},\label{eq:mupq} \\
\mu_k&\coloneqq \mu(p_k,p_{k+1}),\label{eq:mupk} \\
\sigma^2(p,q)&\coloneqq \frac{2\xi^2}{1-\xi}\left(\frac{q^{1-2\xi}-p^{1-2\xi}}{1-2\xi}+p^{1-\xi}\frac{q^{-\xi}-p^{-\xi}}{\xi}+\frac{2p^{1-\xi}q^{1-\xi}-p^{2-2\xi}-q^{2-2\xi}}{2-2\xi}\right)\label{eq:sigpq},\\
\sigma_k^2&\coloneqq \sigma^2(p_k,p_{k+1}), \label{eq:sig2k} \\
\sigma_{jk}&\coloneqq -\xi^2\frac{p_{j+1}^{1-\xi}-p_j^{1-\xi}}{1-\xi}\left(\frac{p_{k+1}^{-\xi}-p_k^{-\xi}}{\xi}+\frac{p_{k+1}^{1-\xi}-p_k^{1-\xi}}{1-\xi}\right),\label{eq:sigjk} \\
\Sigma&=(\Sigma_{jk})\in \R^{L\times L}, \quad \text{where} \quad \Sigma_{jk}\coloneqq \begin{cases}
\sigma_k^2, &(j=k)\\
\sigma_{jk}, & (j<k)\\
\sigma_{kj}, & (j>k)
\end{cases}\label{eq:Sigma}\\
r_k&\coloneqq \mu_k/\mu_L, \label{eq:rk} \\
r&\coloneqq (r_1,\dots,r_{L-1})^\top \in \R^{L-1}, \label{eq:r} \\
H&\coloneqq \begin{bmatrix}
I_{L-1} & -r 
\end{bmatrix}/\mu_L \in \R^{(L-1)\times L},\label{eq:H} \\
\Omega&\coloneqq H\Sigma H^\top, \label{eq:Omega} \\
s_k&\coloneqq \frac{S_{n_k}-S_{n_{k-1}}}{S_{n_{L+1}}-S_{n_L}}, \label{eq:sk}\\
s&\coloneqq (s_2,\dots,s_L)^\top\in \R^{L-1}. \label{eq:s}
\end{align}
\end{subequations}
Here in \eqref{eq:sigpq}, we interpret $\frac{q^{1-2\xi}-p^{1-2\xi}}{1-2\xi}$ as $\log (q/p)$ if $\xi=1/2$.

Let $\alpha_0>1$ be the true Pareto exponent and $r(\alpha)$ be defined as in \eqref{eq:r}. The following theorems summarize the primary results of \cite{TodaWang2021JAE}.

\begin{thm}[Consistency]\label{thm:consistent}
Suppose Assumptions \ref{asmp:iid} and \ref{asmp:F} hold with $t_{L+1}\ge c$. Let $A\subset (1,\infty)$ be compact, $\alpha_0\in A$, and suppose $\hat{W}\pto W$ as $n\to\infty$, where $\hat{W},W$ are symmetric and positive definite. Let $\hat{\alpha}$ be the minimum distance estimator
\begin{equation}
\hat{\alpha}=\argmin_{\alpha\in A}(r(\alpha)-s)^\top \hat{W} (r(\alpha)-s).\label{eq:MDE}
\end{equation}
Then $\hat{\alpha}\pto \alpha_0$ as $n\to\infty$.
\end{thm}

\begin{thm}[Asymptotic normality]\label{thm:asymptotic}
Suppose that the assumptions of Theorem \ref{thm:consistent} hold and $\alpha_0$ is an interior point of $A$. Then $\sqrt{n}(\widehat{\alpha}-\alpha_0)\dto N(0,V)$
as $n\to \infty$, where
\begin{equation}
V=(R^\top WR)^{-1}R^\top W\Omega WR(R^\top WR)^{-1} \label{eq:V}
\end{equation}
for $\Omega=\Omega(\alpha_0)$ and $R=\nabla_\alpha r(\alpha_0)$.
\end{thm}

Proposition 1 of \cite{TodaWang2021JAE} shows that the symmetric matrix $\Omega$ in \eqref{eq:Omega} is positive definite, so $\Omega^{-1}$ is well-defined. The following theorem shows that choosing $\Omega^{-1}$ as the weighting matrix achieves efficiency.

\begin{thm}[Efficiency]\label{thm:efficient}
The asymptotic variance $V$ in \eqref{eq:V} satisfies $V\ge (R^\top \Omega^{-1}R)^{-1}$, with equality when $W=\Omega^{-1}$.
\end{thm}

Theorems \ref{thm:consistent}--\ref{thm:efficient} suggest that the following two step procedure achieves a consistent and asymptotically efficient estimation of $\alpha_0$. First, use any symmetric and positive definite matrix $\hat{W}$ and compute the first stage estimator $\hat{\alpha}_1$ as in \eqref{eq:MDE}. Then, set $\hat{W}=\Omega(\hat{\alpha}_1)^{-1}$ and compute the second stage estimator $\hat{\alpha}_2$ as in \eqref{eq:MDE}. Obviously we can iterate these steps arbitrarily many times and the asymptotic theory remains the same. In our empirical application in Section \ref{sec:results}, we use the initial weighting matrix $\hat{W}=\Omega(2)^{-1}$ (so we set $\alpha=2$, which is a typical value for income) and iterate until $\hat{\alpha}$ converges.

\section{Robustness checks}

\subsection{Definition of capital income}\label{subsec:robust_capital}

\begin{figure}[!htb]
\centering
\includegraphics[width=0.7\linewidth]{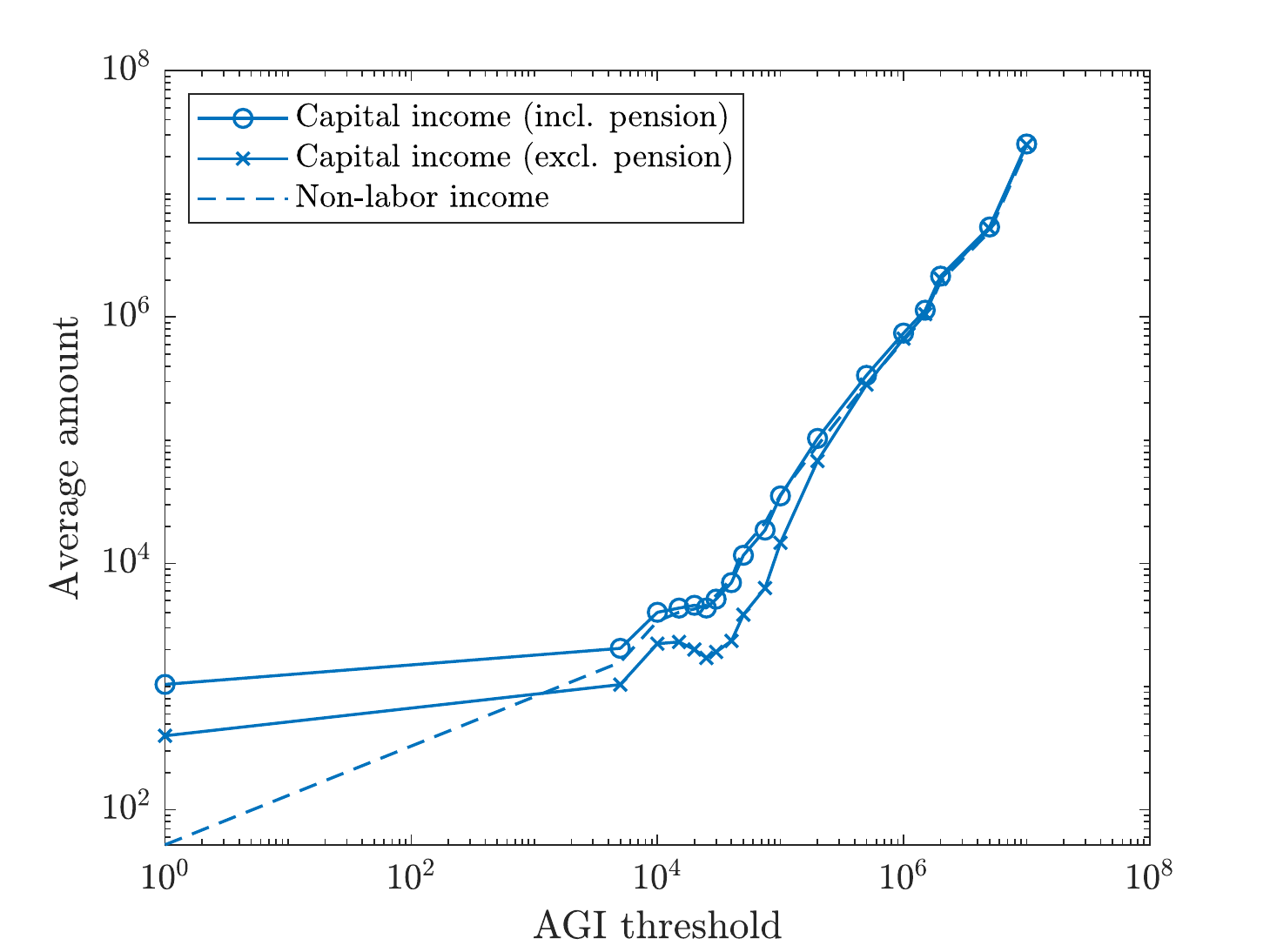}
\caption{Average income of each group, 2019.}\label{fig:average_income2}
\caption*{\footnotesize Note: The figure shows the average income of each income group defined by the lower threshold of AGI. ``Capital income'' is defined by adding up capital income components (including or excluding incomes from IRA distributions, pensions, and annuities). ``Non-labor income'' is AGI minus labor income (salaries and wages).}
\end{figure}

When we defined capital income in Section \ref{subsec:defn_capital} by adding up capital income components, we included ``taxable interest'', ``tax-exempt interest'', ``ordinary dividends'', ``qualified dividends'', ``business or profession'', ``capital gain distributions reported on Form 1040'', ``sales of capital assets reported on Form 1040, Schedule D'', ``sales of property other than capital assets'', ``taxable Individual Retirement Arrangement (IRA) distributions'', ``pensions and annuities'', ``total rent and royalty'', ``partnership and S corporation'', and ``estate and trust''. This definition of capital income is similar to \cite{SaezStancheva2018}, except that they exclude ``taxable Individual Retirement Arrangement (IRA) distributions'' and ``pensions and annuities''. If we follow this narrower definition, Figure \ref{fig:average_income} changes as in Figure \ref{fig:average_income2}. We can see that capital and non-labor income are still nearly identical for high income earners, but the AGI threshold now changes to \$200,000 instead of \$25,000.

\subsection{Definition of sample size}\label{subsec:robust_n}

In Section \ref{subsec:sample_size}, we defined the sample size (the number of tax units) as the adult population ($A$) minus the number of joint returns ($J$), assuming that all married non-filers would file for taxes separately if they were forced to file. Our justification was that it is reasonable to assume that non-filers have low income, and we found that low income earners tend to file separately (Figure \ref{fig:frac_joint_fractile}). Setting the sample size $n=A-J$ is obviously an upper bound. An alternative is to set $n=A-M$, where $M$ is the number of married couples, which is what \cite{PikettySaez2003} use as the sample size. This definition assumes that all married couples file for taxes jointly and gives a lower bound for the number of tax units. A natural concern is whether the definition of the sample size matters for the estimation results.

\begin{figure}[!htb]
\centering
\begin{subfigure}{0.48\linewidth}
\includegraphics[width=\linewidth]{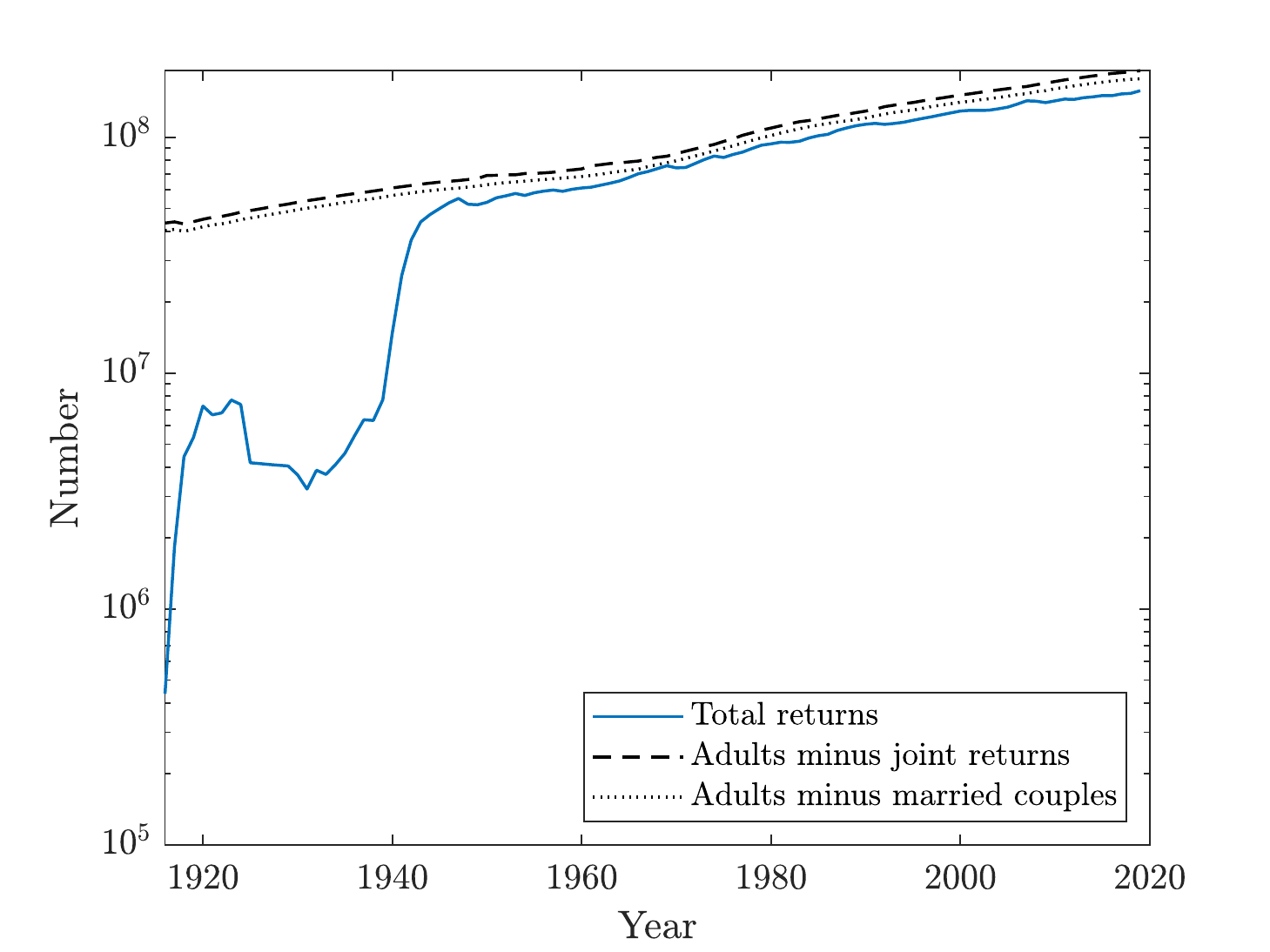}
\caption{Potential tax units.}\label{fig:tax_unit_PS}
\end{subfigure}
\begin{subfigure}{0.48\linewidth}
\includegraphics[width=\linewidth]{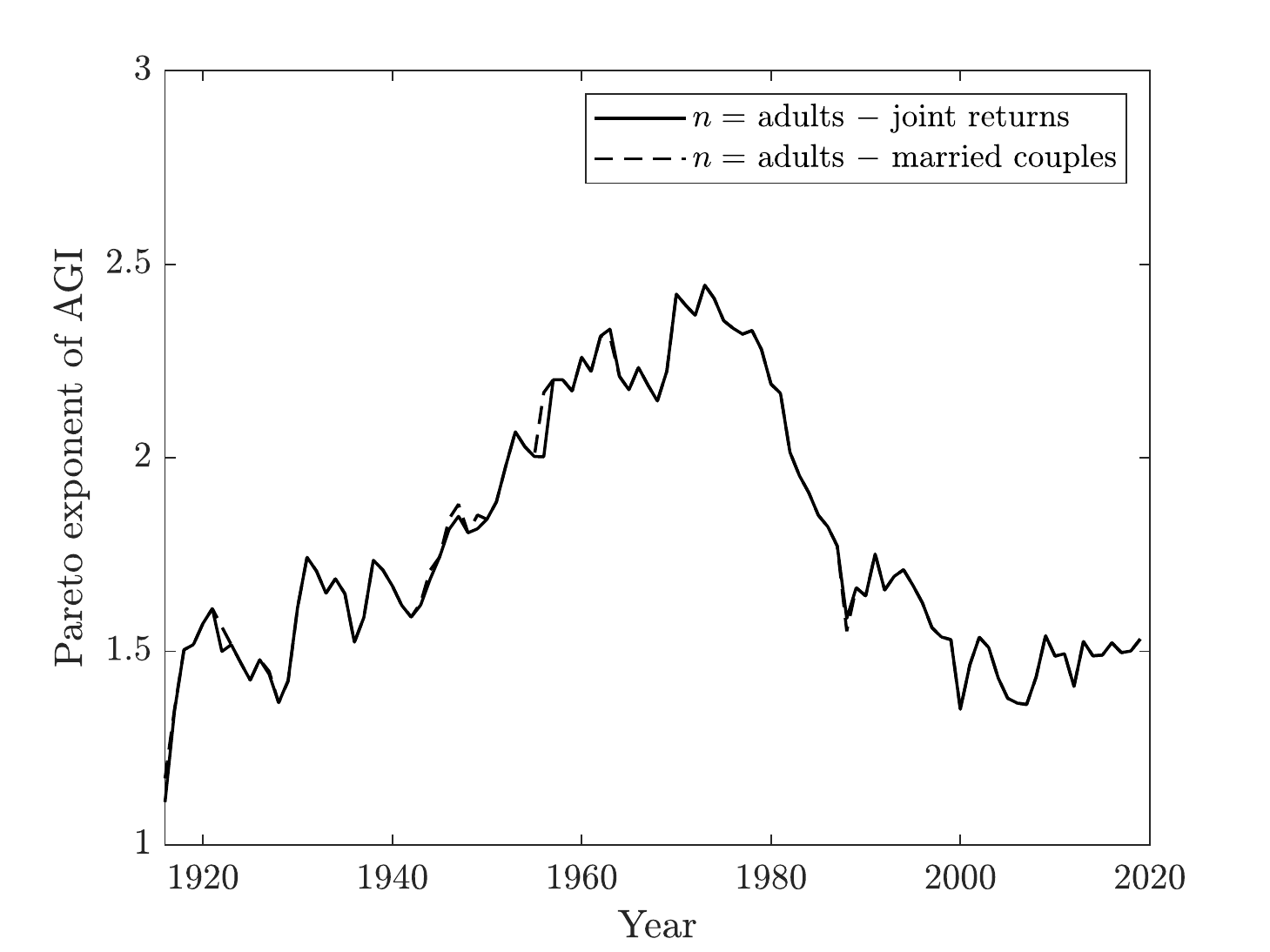}
\caption{AGI Pareto exponents.}\label{fig:alpha_N_PS}
\end{subfigure}
\caption{Sample size and Pareto exponents.}
\end{figure}

Figure \ref{fig:tax_unit_PS} shows the total number of tax returns as well as the two estimates of the number of potential tax units (upper bound $A-J$ and lower bound $A-M$). The two estimates are similar. Figure \ref{fig:alpha_N_PS} shows the estimated AGI Pareto exponents using the two sample sizes. The two estimates are nearly identical, which is natural given the scale invariance of the Pareto distribution. Therefore although we can only bound the number of potential tax units, its exact value has no material impact on the estimation of the Pareto exponents.

\subsection{Choice of income groups}\label{subsec:robust_L}

\begin{figure}[!htb]
\centering
\includegraphics[width=0.7\linewidth]{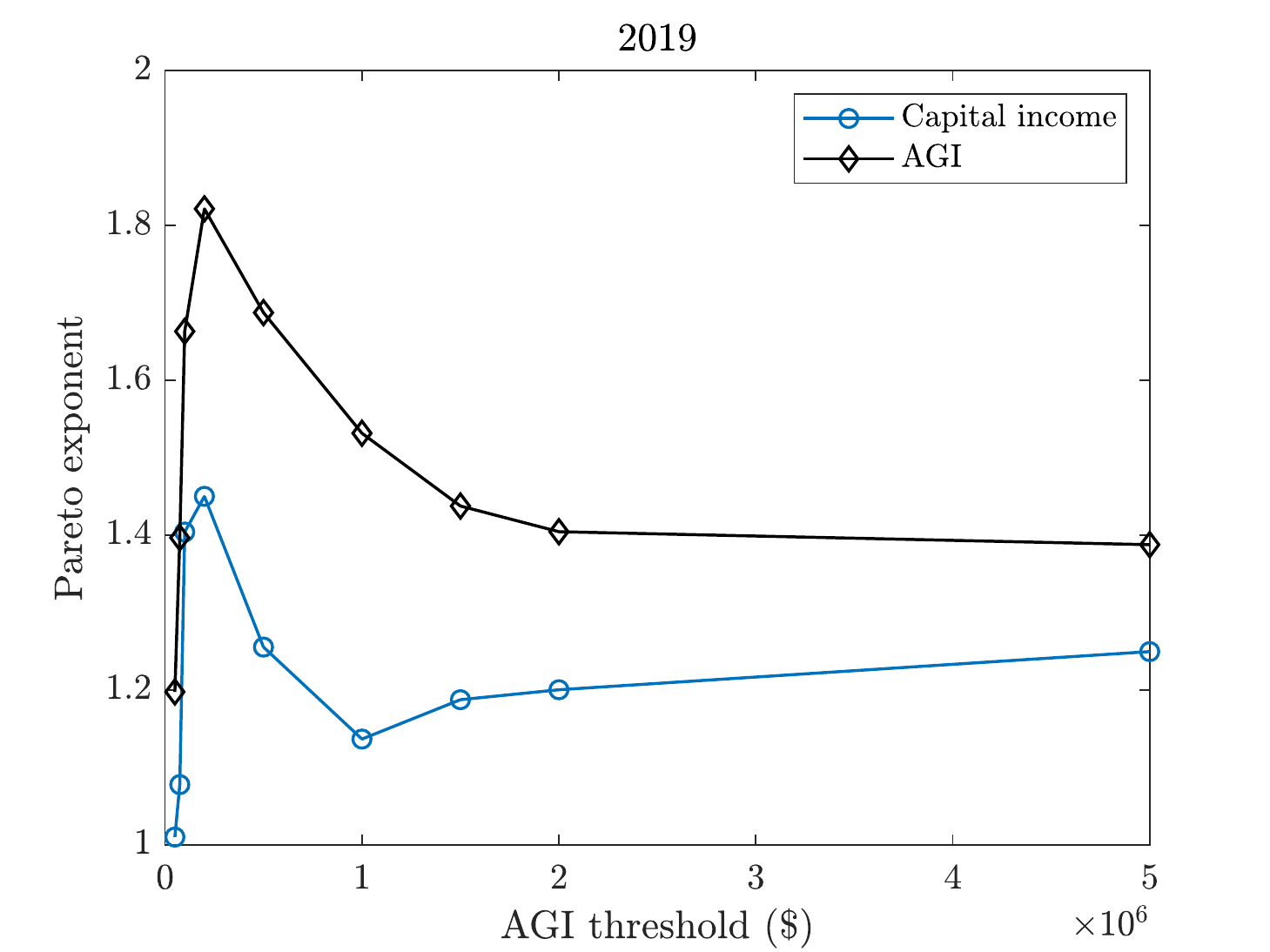}
\caption{Income groups and Pareto exponent.}\label{fig:robust_L}
\caption*{\footnotesize Note: The figure shows the Pareto exponents estimated from income groups above each AGI threshold.}
\end{figure}

In our empirical analysis, we used the top 1\% income earners to estimate the Pareto exponents. Figure \ref{fig:robust_L} shows the dependence of Pareto exponents on the choice of the AGI threshold (the so-called ``Hill plot''). According to Figure \ref{fig:robust_L}, the capital income Pareto exponent estimates are roughly constant beyond AGI threshold of \$500,000, which corresponds to the top 0.9\% income earners. Thus using the top 1\% seems reasonable. On the other hand, the Pareto exponent estimates for AGI are monotonically decreasing in the range \$200,000--\$2,000,000 possibly because AGI is the mixture of capital and labor income, which have different Pareto exponents according to Figures \ref{fig:alpha} and \ref{fig:alpha_labor}.

\subsection{Standard errors}\label{subsec:robust_se}

\begin{figure}[!htb]
\centering
\includegraphics[width=0.7\linewidth]{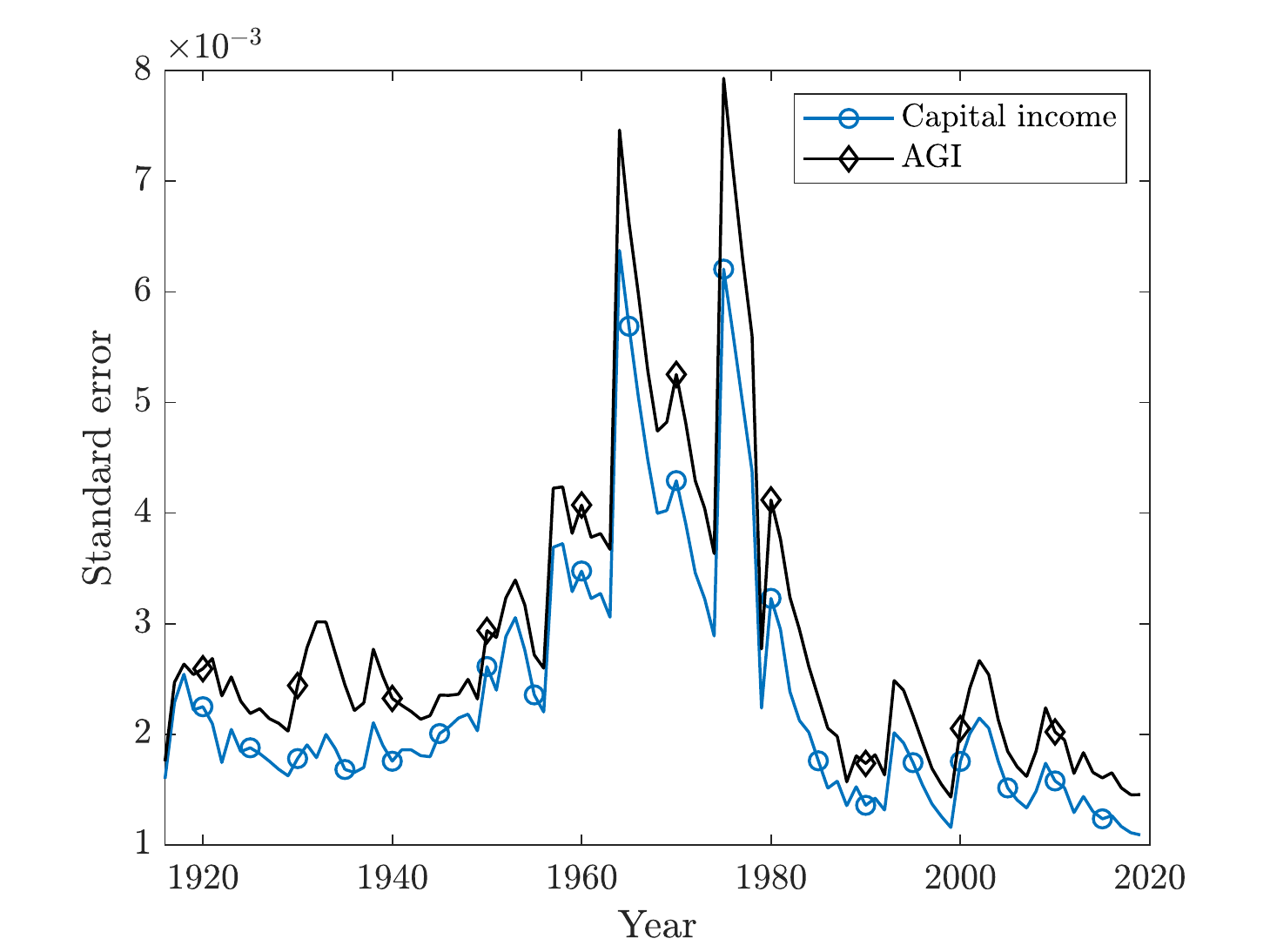}
\caption{Standard errors of Pareto exponents.}\label{fig:alpha_se}
\end{figure}

In Section \ref{sec:results} we omitted standard errors, merely mentioning that they are small. Figure \ref{fig:alpha_se} shows the standard errors of Pareto exponents computed using Theorem \ref{thm:efficient}. The order of magnitude $10^{-3}$ is indeed small.

\subsection{Estimation of AGI Pareto exponents}\label{subsec:robust_estim}

We compare our estimates of Pareto exponents using the method of \cite{TodaWang2021JAE} ($\hat{\alpha}_\mathrm{TW}$) to three existing methods. Because some methods require the income thresholds while others require the partial sums of top incomes, the comparison is possible only for adjusted gross income (AGI), for which all quantities are available. The first method we consider is maximum likelihood ($\hat{\alpha}_\mathrm{ML}$), which is explained in Appendix \ref{subsec:estim_ML}. The second is the method of \cite{FeenbergPoterba1993} ($\hat{\alpha}_\mathrm{FP}$), which is given by \eqref{eq:alpha_FP}. Following their paper, we use the two income thresholds $y_1<y_2$ that bracket the top 0.5\% of tax units. The third is the method of \cite{AtkinsonPiketty2010} ($\hat{\alpha}_\mathrm{AP}$), which is given by \eqref{eq:alpha_AP}. Following their book, we use the top 0.1\% and 1\% income shares constructed by \cite{PikettySaez2003}. As a robustness check, we also use the top income shares constructed ourselves.\footnote{When we construct the top income shares ourselves, we use the total AGI from all tax returns with positive AGI as the denominator, not the number from national accounts used in \cite{PikettySaez2003}. However, this is without loss of generality because the denominator cancels out in the formula \eqref{eq:alpha_AP}. We compute the top 0.1\% and top 1\% income shares by interpolating $\log(\text{top share})$ over $\log(\text{top fractile})$ using spline interpolation, which should be quite accurate because Figure \ref{fig:logshare} shows a straight-line pattern.}

\begin{figure}[!htb]
\centering
\begin{subfigure}{0.48\linewidth}
\includegraphics[width=\linewidth]{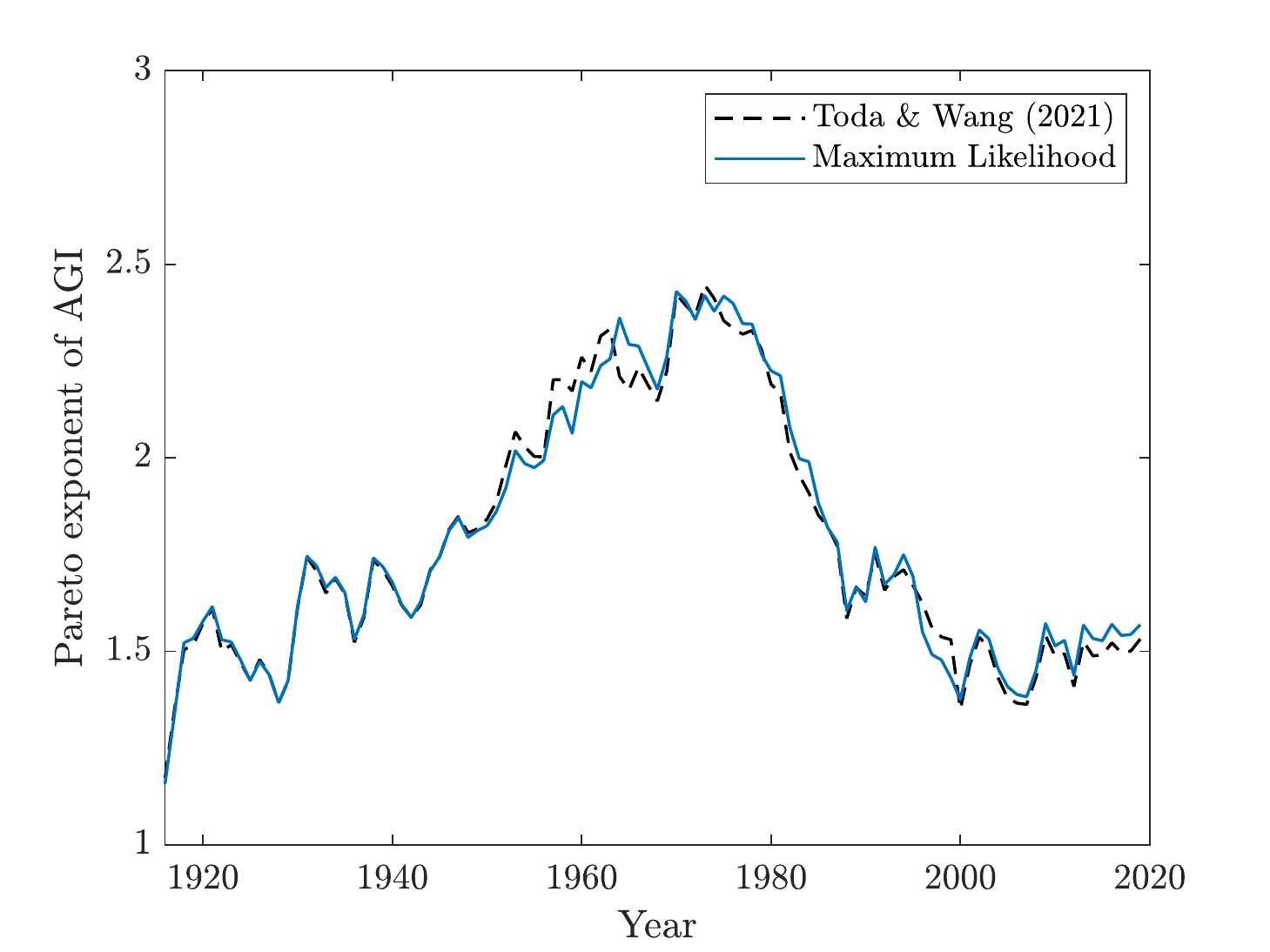}
\caption{Maximum likelihood.}\label{fig:alpha_ML}
\end{subfigure}
\begin{subfigure}{0.48\linewidth}
\includegraphics[width=\linewidth]{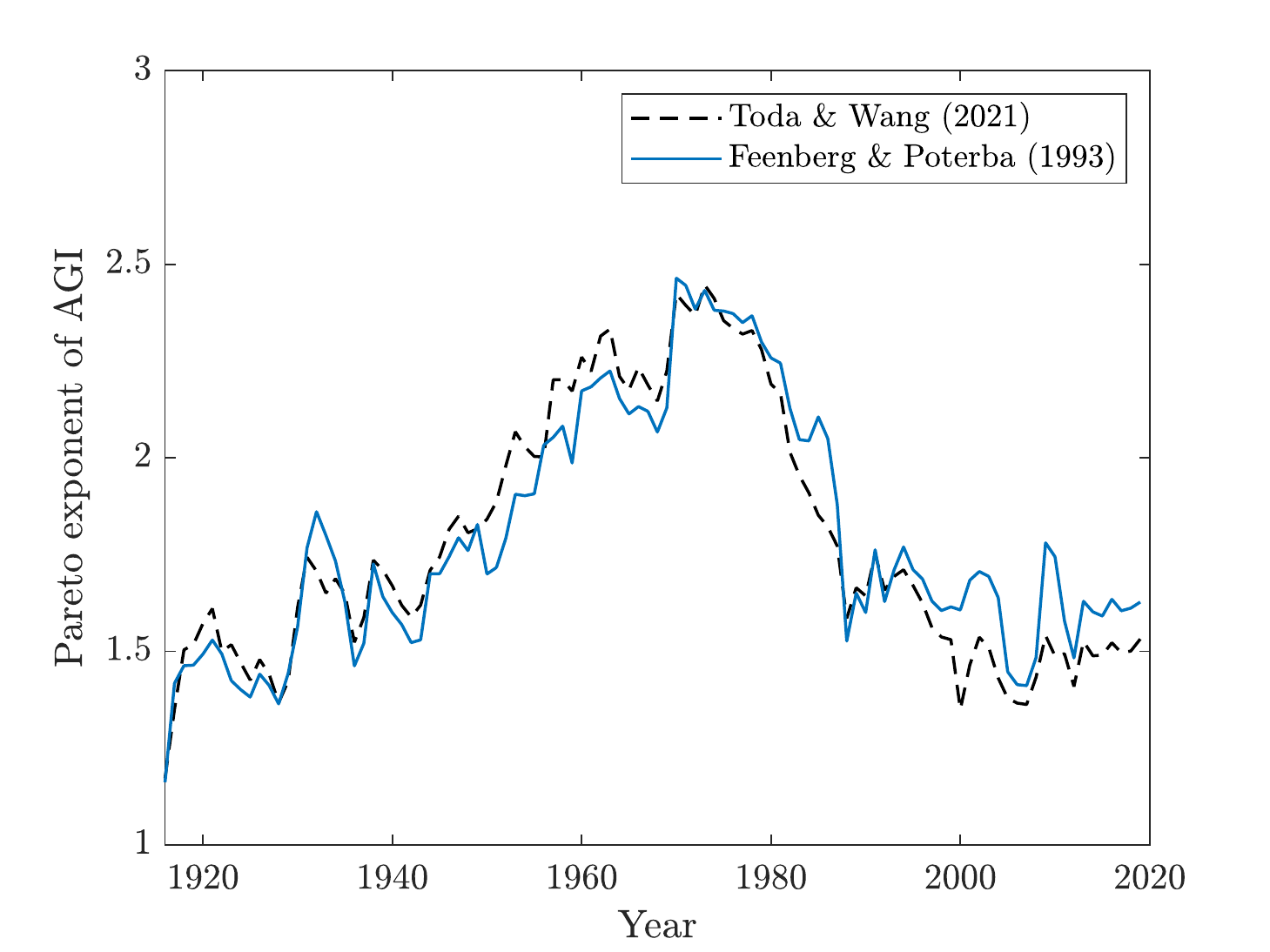}
\caption{\cite{FeenbergPoterba1993}.}\label{fig:alpha_FP}
\end{subfigure}
\begin{subfigure}{0.48\linewidth}
\includegraphics[width=\linewidth]{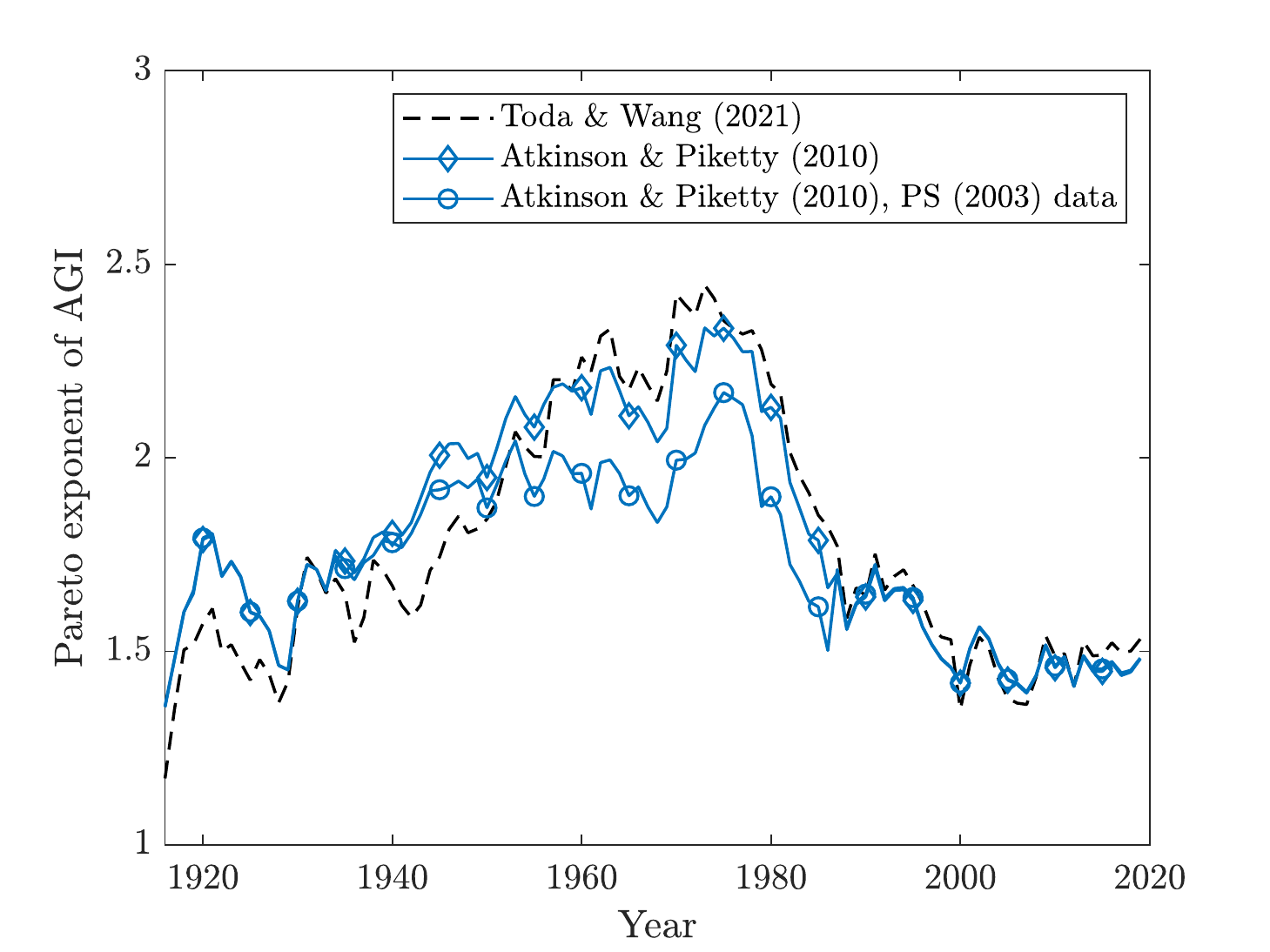}
\caption{\cite{AtkinsonPiketty2010}.}\label{fig:alpha_AP}
\end{subfigure}
\caption{AGI Pareto exponents with different estimation methods.}\label{fig:alpha_AGI}
\end{figure}

Figure \ref{fig:alpha_AGI} shows these estimates. The maximum likelihood estimates $\hat{\alpha}_\mathrm{ML}$ are nearly identical to $\hat{\alpha}_\mathrm{TW}$. This is not surprising because both methods use the full information (either income thresholds or cumulative sums of incomes) and are efficient. The \cite{FeenbergPoterba1993} estimates $\hat{\alpha}_\mathrm{FP}$ are similar to ours but slightly differ, possibly because this method use only the information around the top 0.5\%. The estimates using the \cite{AtkinsonPiketty2010} method tend to be somewhat different. When we apply the estimation methods to the same data sets, $\hat{\alpha}_\mathrm{AP}$ is similar to $\hat{\alpha}_\mathrm{TW}$ post-1950 but larger pre-1950. When we estimate $\hat{\alpha}_\mathrm{AP}$ using the \cite{PikettySaez2003} data set, the estimates are nearly identical to those obtained from our data set pre-1934 and post-1987 but substantially smaller during the period 1935--1986.

\begin{figure}[!htb]
\centering
\includegraphics[width=0.7\linewidth]{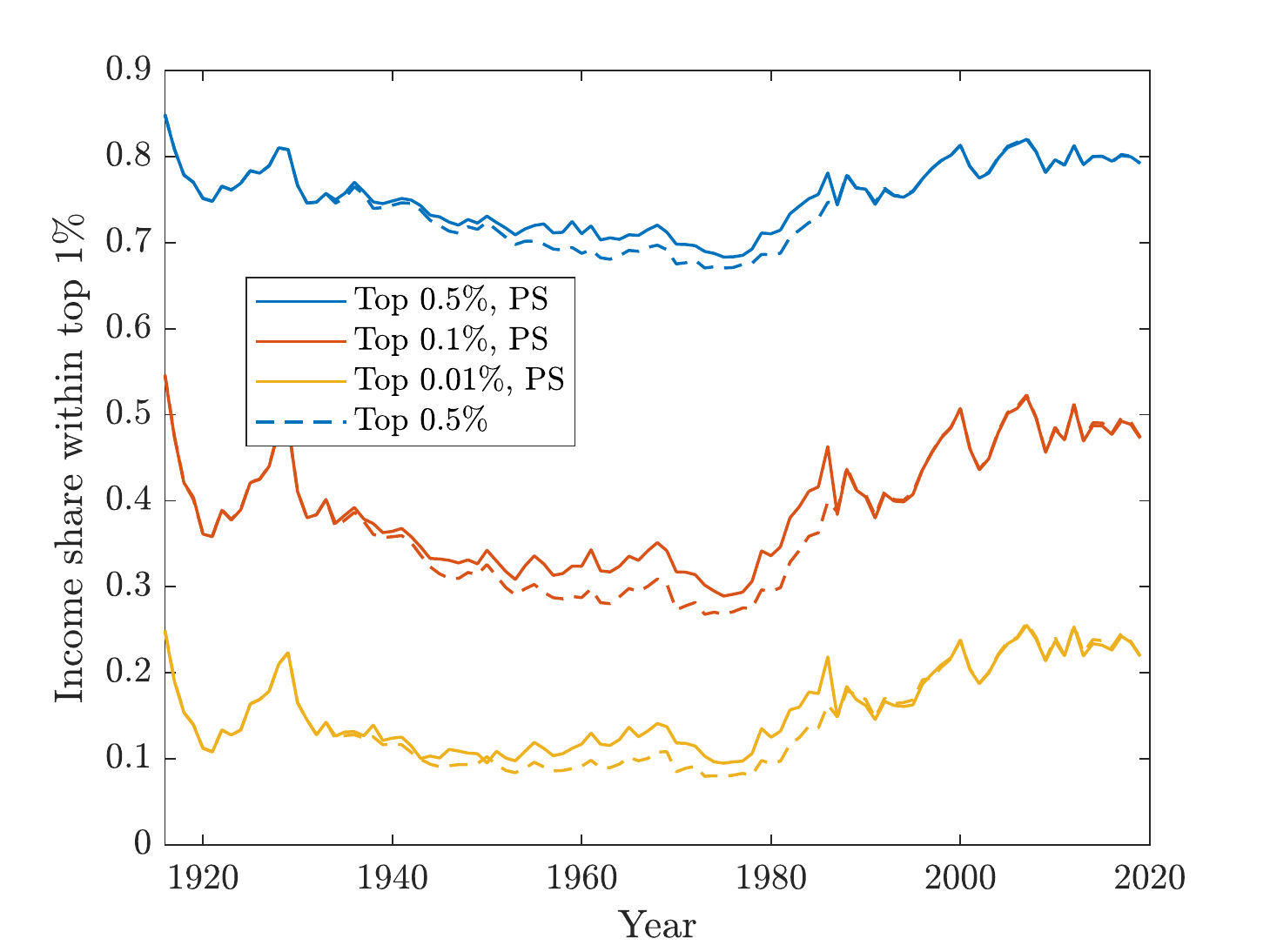}
\caption{Top relative income shares within top 1\%.}\label{fig:relshare}
\end{figure}

To investigate where the difference comes from, Figure \ref{fig:relshare} plots the top relative income shares within the top 1\% income share defined by $S(p)/S(q)$ for $q=0.01$ and $p=0.0001,0.001,0.005$. When the income distribution has a Pareto upper tail, we have $S(p)=Ap^{1-1/\alpha}$ for some constant $A$, so $S(p)/S(q)=(p/q)^{1-1/\alpha}$. Note that this top relative income share depends only on the relative fractile $p/q$, so it is independent of the sample size discussed in Section \ref{subsec:sample_size}. The fact that our relative shares in Figure \ref{fig:relshare} differ from the \cite{PikettySaez2003} relative shares for 1935--1986 suggests that the income concept used in this period is different, although we were not able to identify the exact cause by reading the detailed description in the working paper version \citep{PikettySaez2001WP}.

\end{document}